\documentclass[%
 reprint,
 superscriptaddress,
 amsmath,amssymb,
 aps,
pra,longbibliography]{revtex4-1}
\usepackage[utf8]{inputenc}
\usepackage{hyperref}
\usepackage{graphicx, dcolumn, amsmath, amsfonts}
\usepackage{MySymbols}
\usepackage{upgreek, tikz, dsfont}
\usepackage{subfiles}

\definecolor{dGreen}{RGB}{34,139,34}
\definecolor{dRed}{RGB}{139,34,34}
\definecolor{dBlue}{RGB}{34,34,139}
\definecolor{dGrey1}{RGB}{77,77,77}
\definecolor{dGrey2}{RGB}{123,123,123}
\definecolor{dGrey3}{RGB}{178,178,178}

\begin{document}
\title{Efficient verification and fidelity estimation of discrete bipartite squeezed  states}

\author{Russell P Rundle}
\email{r.rundle@bristol.ac.uk}
\affiliation{School of Mathematics, Fry Building, University of Bristol, UK}

\date{\today}

\begin{abstract}
To gain an advantage, quantum technologies utilize phenomena particular to quantum mechanics. 
Two such phenomena are squeezing and entanglement.
Having generated states that exhibit these features, verification of their generation with local measurements can be a difficult process.
Here we consider the states that are generated using the two-qudit single-axis squeezing Hamiltonian, that not only produces entangled two-qudit squeezed states but also results in various forms of interesting entangled states.
We show how one can use local measurements to both efficiently verify and directly estimate the fidelity of these generated states.
\end{abstract}

\maketitle

\section{Introduction}
Utilizing quantum mechanical effects can lead to advantages in quantum technologies over their classical counterparts.
It is therefore important to devise methods to effectively and efficiently verify quantum processes and states.
One could choose to reconstruct the quantum state by means of quantum state tomography~\cite{QuantumTomography,James2001,DAriano2002,Gross2010}.
However, as the size of the quantum system in question increases, it becomes increasingly more difficult and time-consuming to fully reconstruct a quantum state.
It then becomes more important to find a more computationally-efficient method to verify that a desired state has been created; and to verify that a given computation or procedure was performed correctly.

There have been many alternative approaches to verify the generation of a quantum state~\cite{Kliesch2020, GheorghiuVerification, EisertVerifiction}.
These include considering and measuring certain properties and attributes via an appropriate metric; this has been done for properties such as entanglement~\cite{Guhne2009,HorodeckiHorodeckiHorodeckiHorodecki}, entropy, purity, coherence~\cite{Baumgratz2014,Streltsov2015}, among others, see \Refe{MontanaroDeWolf} for a survey of quantum property testing.
Alternatively methods to more directly certify the creation of a state, such as estimating the fidelity~\cite{FlammiaLiu, daSilva2011} and quantum state verification~\cite{Pallister, Masahito2015, ZhuHayashi2019, LiZihao2019, WangKun2019, YuXiaoDong2019}, have been devised recently.

In quantum state verification the goal is to devise a strategy that is made up of a set of measurements, where each measurement accepts a given target state with certainty and rejects any other state with some probability~\cite{Pallister}.
The strategy is then a convex sum of these measurements, in a way that optimizes the rejection of a state that isn't the target state.
Measurements are then taken in line with this strategy a given number of times.
If any of these measurements fail, we can say with certainty that the generated state is not the target state.

This method of verification assumes that the target state is pure, however in experimental situations sufficient purity often proves difficult to achieve.
Resulting in a non-zero chance that the generated state is not the target state.
It may then be more useful to instead accept that the generated state is not the target state, and calculate how close you are to generating the state, with respect to some distance measure in state space.
This is where protocols such as direct fidelity estimation are useful~\cite{FlammiaLiu,daSilva2011}.
By taking a sequence of measurements, one can calculate the fidelity by using the measured results with the theoretical results of the target state.

Here, we extend previous results for quantum state verification and fidelity estimation to bipartite-qudit systems. 
As the dimension of qudits increase, we then need $d-1$ degrees of freedom to uniquely define a bipartite qudit state, up to local rotations; we therefore consider how states that are generated by a two-qudit squeezing Hamiltonian can be verified.
This will require just a time degree of freedom and allows us to consider particular states that may be of interest.

We will first consider an extension to the quantum state verification protocol found in \Refe{Pallister}, where we will discuss the previous results for two-qubit systems and then extend this to larger discrete systems.
This will be followed by how one can extend the direct fidelity estimation method from \Refe{FlammiaLiu} to bipartite qudit systems.

\section{The Model: Squeezing and entanglement}
Squeezed states of light are generated by squeezing a coherent state of light along a given quadrature.
Such states have been demonstrated to give an advantage in various applications~\cite{Breitenbach}, such as in continuous-variable quantum computing and quantum information processing~\cite{Braunstein2005,Furusawa1998,Menicucci2008}. 
They have also been useful in quantum metrology~\cite{Polzik1992,Lawrie2019}, where squeezed states of light are used in the detection of gravitational waves~\cite{Chua2014,LIGOSqueezed1,LIGOSqueezed2}.
Alternatively, given two modes of light, one can squeeze the state along quadratures shared by the two modes, this creates an entangled state across these two modes.
When considering optical systems restricted to Gaussian states, one way to generate entanglement between two modes is to apply this squeezing operation across the two modes~\cite{Braunstein2005}.

Conversely, there has been considerable attention into the generation of squeezed states in finite Hilbert spaces. 
Where spin squeezed states have proven to be a valuable resource in the generation of atomic clocks~\cite{Leroux2010,Louchet-Chauvet2010}.
One method to produce such states is by applying the single-axis twisting Hamiltonian~\cite{Agarwal1997}.
Further application of this Hamiltonian then produces atomic Schr\"odinger cat states -- the superposition of many spin coherent states~\cite{Agarwal1997}.
By considering a single large spin as the symmetric subspace of many qubits, the single-twisting Hamiltonian has also been a reliable method to generate GHZ states of up to 20 qubits~\cite{Song2019}.

Analogously to a two-mode squeezed state, one can adapt the single-axis twisting Hamiltonian to generate a two-qudit squeezing operation.
Given two spin-$j$ qudits, each with dimension $d = 2j+1$, two-qudit squeezed states can be generated by applying the Hamiltonian
\begin{equation}\label{TwoQuditTwisting}
	H_d = \mu \OpJz_d  \otimes \OpJz_d,
\end{equation}
where $\OpJz_d$ is the $d$-dimensional spin operator that rotates a $d$-dimensional state around the $z$-axis.

This Hamiltonian can then be applied to an initial state $\ket{\Psi(0)}_d$, where
\begin{equation}\label{intialState}
	\ket{\Psi(0)}_d = \frac{1}{2^{d-1}}\,\sum_{k=0}^{d-1} \sum_{k'=0}^{d-1} \left[\binom{d-1}{k}\binom{d-1}{k'}\right]^{\frac12}\ket{k}\ket{k'}.
\end{equation}
This state is the eigenstate of the $\OpJx_d \otimes \OpJx_d$ operator with eigenvalue $(d-1)^2/4$, generating a spin-coherent state that is aligned along the equator of the Bloch sphere~\cite{Arecchi1972,PerelomovB}.
For two qubits, this state is simply $(\ket{0}+\ket{1})\otimes (\ket{0}+\ket{1})/2$.

Applying $\exp\left(-\ui \tau H_d\right)$ to \Eq{intialState}, where $\tau = \mu t$ is dimensionless time, we yield the time-dependent state
\begin{eqnarray}\label{GeneralTimeDepState}
	\ket{\Psi(t)} &=& \ue{-\ui \tau \OpJz_d  \otimes \OpJz_d} \ket{\Psi(0)}_d \\
	& =& \sum_{k,k'=0}^{d-1} \frac{\ue{-\ui\tau(j-k)(j-k')}}{2^{d-1}} \left[\binom{d-1}{k}\binom{d-1}{k'}\right]^{\frac12}\ket{k}\ket{k'}. \nonumber
\end{eqnarray}

\begin{figure}
	\includegraphics[width = \linewidth]{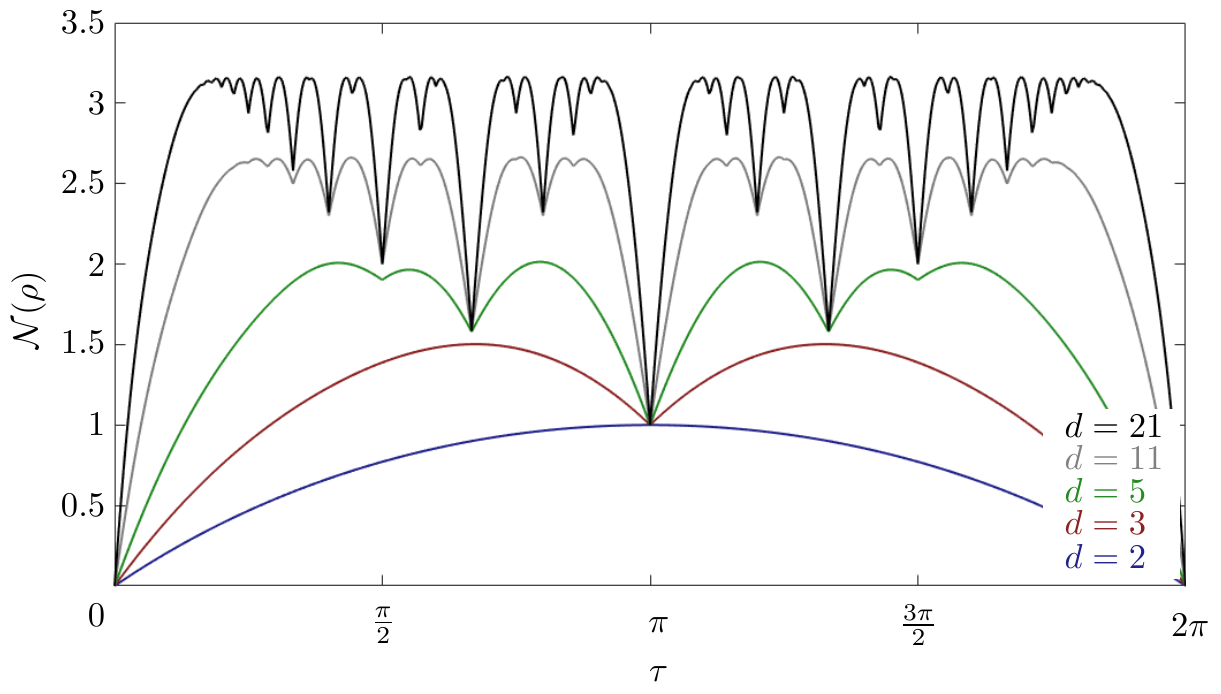}
	\caption{\label{Negativity}
	The negativity of bipartite qudit states through the two-qudit squeezing evolution.
	The blue curve (bottom) is the negativity of two qubits; the red (middle) line is the negativity of two qutrits; the following green curve is for two 5-level qudits.
	The top to curves show the values for two 11-level and 21-level qubits in grey and black respectively.
	This quantifies the entanglement between the two qudits at different times in the evolution.
	At each of the troughs in the negativity particularly interesting entangled states are generated, and will be considered separately in \Sec{specialCases}.}
\end{figure}

The evolution of this initial state under this Hamiltonian results in the state twisting around the $z$ axis, causing the initial coherent state to squeeze, creating a discrete analog of two-mode squeezing as the two states entangle~\cite{ChanLaw2006}.
This has also been considered as a method to entangle two Bose Einstein condensates~\cite{Byrnes2012,Byrnes2013,MengXin2021}, to treat Bose Einstein condensates as qubits in quantum computations.
A method to verify the creation of such states can therefore be used to verify the generation of entangled Bose Einstein condensates.

Given we are only considering pure states during the evolution, one can use multiple metrics to consider the entanglement, such as the purity~\cite{ChanLaw2006} or von Neumann entropy.
Here we consider the negativity, where
\begin{equation}
	\mathcal{N}(\rho) = \frac{||\rho^{\Gamma_A}||_1-1}{2},
\end{equation}
$\rho^{\Gamma_A}$ is the partial transpose of the density matrix $\rho$ with respect to qubit $A$.
$||X||_1=\Trace{\sqrt{X^\dag X}}$ is the trace norm.

From \Fig{Negativity}, it can be seen that as the squeezing occurs, the negativity between the two systems increases.
Unlike the infinite-dimensional case, different forms of entangled states arise throughout the evolution.
These can be identified by the dips in the negativity. 
A curious effect is the fractal-like behaviour throughout the evolution~\cite{Byrnes2013}.

At $\tau=\pi$, the negativity is always equal to 1, regardless of the dimension.
In all cases, this state is of the same form; that is, locally equivalent to $(\ket{00}+\ket{11})/\sqrt{2}$, and for two qubits is a maximally entangled state.
Alternatively, we can think of it as a bipartite-qubit, two-component atomic Schr\"odinger cat state. 
Such a state will be labelled $\cats{2}{2}$ where the first 2 represents the dimension of each qudit and the second 2 represents the number of components.

As the dimension increases, the form of the state is the same, albeit with a larger overall Hilbert space, where we will label a general state in the Schmidt decomposition at $\tau=\pi$ as
\begin{equation}\label{TwoCompCat}
	\cats{d}{2} = \frac{1}{\sqrt{2}} (\ket{00}_d + \ket{11}_d),
\end{equation}
where $d$ is the dimension of each qudit.

When the dimension increases, other forms of entanglement start to appear.
These can be seen in the local minima in the negativity for higher-dimensional systems.
The states at these points are bipartite-qudit, $\kappa$-component atomic Schr\"odinger cat states. 
In the Schmidt decomposition, these states can be written as
\begin{equation}\label{nCompCat}
	\cats{d}{\kappa} = \frac{1}{\sqrt{\kappa}} \sum_{k=0}^{\kappa-1} \ket{kk}_d.
\end{equation}
Note that when $\kappa = d$ these state are maximally entangled states.
Alternatively, when  $\kappa < d$ the state is maximally in a subspace of the full Hilbert space, however the full state itself is not maximally entangled.

Note that only the two-qubit state creates a maximally entangled state.
Interestingly, when the dimension is large enough, the state $\cats{d}{\kappa}$ appears at $\tau = 2\pi/\kappa$, and certain multiples thereafter -- unless coinciding with a lower value of $\kappa$.
For example a $\kappa = 4$ state appears at $\tau = \pi/2$ and $3\pi/2$, but not at $\tau = \pi$ where a $\kappa = 2$ state appears.

All the entangled states that appear at these dips in negativity are special cases of \Eq{GeneralTimeDepState} that have a more simple procedure for optimal verification; these cases will be considered separately in \Sec{VerificationSpecialCases}.
Following this, we will consider the methods that we can verify any general state of the form \Eq{GeneralTimeDepState}.

\section{Existing Verification Methods}

There are cases where verification methods have already been given.
In \Refe{Pallister}, a method to optimally verify any two-qubit state was presented, where the authors also presented a method to verify stabilizer states.
This was followed results in \Refe{ZhuHayashi2019} where any multi-qubit maximally entangled state can be generated.
We also note that adaptive protocols have been presented both for the two-qubit case and for two-qudits~\cite{LiZihao2019,YuXiaoDong2019}, where such protocols require communication between measurements. 
Here we are interested in generalizing a non-adaptive protocol to any two-qudit system.
The results of this will be given in \Sec{GeneralizedSrategies}.
First, we will discuss the two-qubit case from \Refe{Pallister}, along with the maximally entangled states considered in \Refe{ZhuHayashi2019} and a discussion around certain types of special cases.

\subsection{Special cases}\label{specialCases}\label{VerificationSpecialCases}

Verification of bipartite states with local measurements can be complicated procedure, where there has been much work on providing methods to do verify arbitrary quantum states~\cite{Pallister, Masahito2015, ZhuHayashi2019, LiZihao2019, WangKun2019, YuXiaoDong2019}.
There are however certain states that require a much simpler procedure than the general case, it is then worth considering these cases first.

An important class of these are maximally entangled states
\begin{equation}\label{maxEntState}
	\cats{d}{d} = \frac{1}{\sqrt{d}}\sum_{k=0}^{d-1} \ket{kk}_d.
\end{equation}
It was shown in \Refe{ZhuHayashi2019} such states for any dimension can be verified.
Where for prime-dimensional systems, the strategy can be simply calculated with the Weyl algebra, otherwise a strategy can be generated through a 2-design structure~\cite{ZhuHayashi2019}.

For two qubits, the verification of the maximally entangled state $\ket{\psi} = (\ket{00}_2+\ket{11}_2)/\sqrt{2}$ can be performed by the strategy
\begin{equation}\label{BellStrategyTwoQubits}
	\Omega_{\text{Bell}} = \frac{1}{3} \left(\Pi_2^1+\Pi_2^2+\Pi_2^3\right),
\end{equation}
made up of the projectors $\Pi_2^i$, such that each
\begin{equation} \label{TwoQubitBellProjectors}
	\Pi_2^i = \sum_{k\in\{-1,1\}}\Proj{\sigma_i,k}_2 \otimes {\Proj{\overline{\sigma_i,k}}}_2
\end{equation}
where the subscript refers to $d=2$. 
The states in \Eq{TwoQubitBellProjectors} are the eigenstates of the Pauli operators $\sigma_i\ket{\sigma_i,k}_2 = k \ket{\sigma_i,k}_2$ and $\ket{\overline{\sigma_i,k}}_2$ is the complex conjugate of the vector.
Note that every two-qubit maximally entangled state is simply a local rotation of this state.

When considering the evolution of \Eq{TwoQuditTwisting}, at $t=\pi$ a state of this form is produced regardless the dimension of the qudits, producing a state of the form $\ket{(d,2)}$ in \Eq{TwoCompCat}.
When $d>2$, $\ket{(d,2)}$ is not maximally entangled. 
This can be seen in comparison to the the two-qutrit maximally entangled state calculated from \Eq{maxEntState}. 
Further we can see this by the value of the negativity of this state in \Fig{Negativity}.
These are still interesting entangled states, it is therefore worthwhile discussing how such states can be verified. 
 
To verify $\ket{(d,2)}$, consider the state $(\ket{00}_d+\ket{11}_d)/\sqrt{2}$, that is equivalent to the target state up to local rotations.
This state is then an eigenstate of the first three generators of the \su{d} algebra. 
The matrix representation of these three can be given by
\begin{equation}
\Lambda_d^i  =	 \begin{pmatrix} \sigma_i  & \vline & 0 \\ \hline 0 & \vline &  0_{d-2} \end{pmatrix},
\end{equation}
where $i=1,2,3$.
These are $d\times d$ matrices with with the Pauli operators in the top-left.

Similarly to \Eq{TwoQubitBellProjectors}, we can define the projectors
\begin{equation}
	\Pi_d^i = \sum_{k\in\{-1,1\}}\Proj{\Lambda_d^i,k} \otimes {\Proj{\overline{\Lambda_d^i,k}}}
\end{equation}
where $\Lambda_d^i\ket{\Lambda_d^i,k} = k \ket{\Lambda_d^i,k} $ and $\ket{\overline{\Lambda_d^i,k}}$ is the complex conjugate of the vector.
The strategy for $\ket{(d,2)}$ can then be built by
\begin{equation}\label{BellGenStrategy}
	\Omega_{\text{Bell}} = \Pi_d^1+\Pi_d^2+\Pi_d^3,
\end{equation}
This can be generalized to measure any entangled state of the form in \Eq{nCompCat}.
Where just a $\kappa$-dimensional subspace of $\ket{(d,\kappa)}$ needs to be measured.

Note in the two-qudit squeezing evolution, that as the dimension of the qudits increases, more variants of these states emerge.
Each of the troughs in the negativity in \Fig{Negativity} is one of these entangled states. 
We can treat these states as subspaces of the full two-qudit space, much as we did in \Eq{BellGenStrategy}.

We now have a method to efficiently verify each of the local minima in the negativity.
However this misses out an important part of the evolution, the spin-squeezed states that are produced at the beginning.

\subsection{Two-qubit verification}

The optimal method to verify any two-qubit state was given in \Refe{Pallister}; we will now present the main results from \Refe{Pallister} in order to consider a method to scale the generated strategy up to larger system sizes.
Note that other strategies have been considered that are adaptive~\cite{LiZihao2019,YuXiaoDong2019}, here our goal is to generalize the non-adaptive strategy in \Refe{Pallister} to larger qudits to create a protocol that is efficient and easily scalable.

We start with a general two-qubit state as our target state.
The general target state given in \Refe{Pallister} is the same as the Schmidt decomposition of two-qubit solution of \Eq{GeneralTimeDepState} 
\begin{equation}\label{twoQubitState}
	\ket{\psi(\tau)} = \cos \tau \ket{00} + \sin \tau \ket{11},
\end{equation}
where the only difference with the target state in \Refe{Pallister} is that the $\sin$ and $\cos$ terms are swapped.

In \Refe{Pallister}, the authors started with a general strategy for this state  that is made up of the convex sum
\begin{equation}\label{GenStrategy2Qubits}
	\Omega = c_1 \Omega_1 + c_2 \Omega_2 + c_3 \Omega_3 + c_4 \Omega_4,
\end{equation}
where $\Omega_i$ is some rank-$i$ strategy.
At $\tau=0$ and $\tau = 2\pi$ all that is needed is a rank-1 strategy for optimal verification.
Since such states are themselves separable, the strategy will simply be $\Proj{\psi}$.
As was shown in \Sec{specialCases}, there is also a special case when $\tau = \pi$, where we only need the rank-2 strategy given in \Eq{BellStrategyTwoQubits}

It was shown in \Refe{Pallister}, that for any other state, we need the addition of a rank-3 strategy as well as the rank-2 operator
\begin{equation}
	\Pz{2} = \Proj{00}_2+\Proj{11}_2.
\end{equation}
The required rank-3 strategy is of the form
\begin{equation}\label{genQubitRank3Strat}
	\Omega_3 = \Bid - \sum_j \eta_j \Proj{\phi_j}
\end{equation}
for some normalization $\eta_j$, where 
\begin{equation}\label{twoQubitOrthogonality}
	\IP{\phi_j}{\psi(\tau)} = 0
\end{equation} 
for every $j$ and each $\ket{\phi_j}$ is separable. 
This results in the overall strategy
\begin{equation}\label{OptimalSolutionTwoQubitsProto}
	\Omega = \alpha \Pz{2}+(1-\alpha)\Omega_3.
\end{equation}

Now all that is needed is the explicit form of \Eq{genQubitRank3Strat} and the value of $\alpha$.
In \Refe{Pallister}, the authors proved that the optimal form of \Eq{genQubitRank3Strat} is given when $\eta_j = 1/3$ for $j=1,2,3$, and 
\begin{equation}
\begin{split}\label{twoQubitOrthStates}
	\ket{\phi_j} =& \left( \frac{1}{\sqrt{1+\cot\tau}}\ket{0} + \frac{\ue{\frac{2\ui\pi j}{3}}}{\sqrt{1+\tan\tau}}\ket{1} \right) \\
	& \otimes \left( \frac{1}{\sqrt{1+\cot\tau}}\ket{0} + \frac{\ue{\frac{2\ui\pi (3-2j)}{3}}}{\sqrt{1+\tan\tau}}\ket{1} \right)
\end{split}
\end{equation}
note that the coefficients are swapped here in comparison to \Refe{Pallister}.
Also note that only three values of $j$ are needed in the sum, this is equivalent to integrating over all valid values of the phases on the states.

The phase difference on the first and second qubit is fixed, this can be seen by considering the requirement in \Eq{twoQubitOrthogonality}. 
This requirement means that any state $\ket{\phi_j}$ needs to have a value $\nu \sin\tau$ as a coefficient of $\ket{00}$ and $-\nu \cos\tau$ as a coefficient of $\ket{11}$, for some normalization $\nu$. 
This results in $\IP{\phi_j}{\psi(\tau)} = \nu \sin\tau \cos\tau - \nu \cos\tau\sin\tau = 0$.

One way to arrive at the states in \Eq{twoQubitOrthStates} is then to start with a state 
\begin{equation}
\begin{split}\label{twoQubitStartOrthStates}
	\ket{\phi_j} =& \nu\left( \sqrt{\sin\tau}\ket{0} + \ue{\ui\varphi}\sqrt{\cos\tau}\ket{1} \right) \\
	& \otimes \left( \sqrt{\sin\tau}\ket{0} + \ue{\ui(\pi-\varphi)}\sqrt{\cos\tau}\ket{1} \right).
\end{split}
\end{equation}
Normalizing results in $\nu = 1/\sqrt{1 + 2\sin\tau\cos\tau}$. 
This gives the correct values in \Eq{twoQubitOrthStates}.
The next step is to integrate over the values of $\varphi$, it was shown in \Refe{Pallister} that taking the sum over the three phases in \Eq{twoQubitOrthStates} yields equivalent results to integrating over all phases.

Now that both parts of the strategy are generated, it is necessary to find the optimal value of $\alpha$.
This means optimizing between the $\Pz{2}$ and the $\Omega_3$ components of the strategy.
Note that $\Pz{2}$ accepts both the target state and the orthogonal state $\ket{\psi^\perp(\tau)} = \sin\tau\ket{00} + \cos\tau\ket{11}$ with certainty while also always rejecting $\ket{01}$ and $\ket{10}$.
$\Omega_3$, on the other hand, accepts the target state with certainty and rejects $\ket{\psi^\perp(\tau)}$ with some given probability; $\Omega_3$ also accepts $\ket{01}$ and $\ket{10}$ with some probability.

Optimizing for $\alpha$ is then a balance between these two scenarios, where we want to decrease the probability of accepting $\ket{\psi^\perp(\tau)}$ as much as possible while not increasing the probability of accepting $\ket{01}$ and $\ket{10}$ too much.
This results in solving $\bra{\psi^\perp(\tau)} \Omega \ket{\psi^\perp(\tau)} = \bra{01} \Omega \ket{01}=\bra{10} \Omega \ket{10}$, resulting in a value of $\alpha= (2-\sin 2\tau)/(4+\sin 2\tau)$.
For more detail into the proof of this optimization, see the supplementary material of \Refe{Pallister}.

\subsection{Two-qubit fidelity estimation}

In experimental settings it is known that the state created isn't exactly the target state.
It can then be much more desirable to estimate the fidelity of the created state with respect to the target state.

This can be done in various ways, one of which is an extension of the two-qubit verification strategy, where it was shown in \Refe{ZhuHayashi2019-2} that the infidelity is bounded by the second highest and the lowest eigenvalue of the strategy.
These inequalities are then saturated when the strategy is of a homogenous form -- when these eigenvalues are equal.

Here we will instead focus on the direct fidelity estimation protocol laid out in \Refs{FlammiaLiu,daSilva2011}.
In \Refe{FlammiaLiu} a protocol to estimate the fidelity of multi-qubit states and processes was given by considering the characteristic function of the state. 
Similar results were shown in \Refe{daSilva2011} where they also extended their results to continuous-variable systems.
In \Refe{daSilva2011}, the authors considered the use of the Wigner function for a continuous-variable system, which is the Fourier transform of the characteristic function.
We note that any informationally complete probability distribution function can be used to perform the procedure for any system, the procedure just may need some adjustment~\cite{Kliesch2020, Rundle2021}.

For estimating the fidelity of a general two-qubit state, we begin with introducing the two-qubit characteristic function, where
\begin{equation}\label{TwoQubitCharacteristic}
	\cf_\rho(k,k') = \frac{1}{2}\Trace{\rho\;\sigma_k\otimes\sigma_{k'}}.
\end{equation}
Note that the characteristic function is normalized so that $\sum_{k,k'}\chi(k,k')^2=1$.
For our two-qubit state in \Eq{twoQubitState} the characteristic function is only non-zero in six elements
\begin{eqnarray}
	\chi(0,0) = \chi(3,3)  &=& \frac{1}{2}\nonumber\\
	\chi(1,1) = -\chi(2,2) &=& \frac{1}{2}\sin2\tau \label{targetCharacteristic} \\
	\chi(0,4) =  \chi(4,0) &=& \frac{1}{2}\cos2\tau.\nonumber
\end{eqnarray}
This means we only need to take measurements in six bases where, from this, one can then calculate the fidelity
\begin{equation}
	F(\rho_1,\rho_2) = \sum_{k,k'} \chi_{\rho_1}(k,k')\,\chi_{\rho_2}(k,k'),
\end{equation} 
where $\chi_{\rho_1}(k,k')$ is the characteristic function for the target state in \Eq{targetCharacteristic} and $\chi_{\rho_2}(k,k')$ is the measured characteristic function.

Following the protocol in \Refs{FlammiaLiu,daSilva2011}, we choose 
\begin{equation}\label{theEll}
	\ell = \lceil 1/\epsilon^2\delta \rceil
\end{equation}
values of $(k,k')$, where $\epsilon$ is the adaptive error and $\delta$ is the failure probability~\cite{FlammiaLiu,daSilva2011,Kliesch2020}.
Each sample will be chosen with the probability $|\cf(k_i,k'_i)|^2$.
On average, the measurements $\Bid\otimes\Bid$ and $\sigma_z\otimes\sigma_z$ will be selected $25\%$ of the $\ell$ times each, and together take up half the basis choices for measurement.
The other half come from the other four measurement bases; where on average $\sigma_x\otimes\sigma_x$ and $\sigma_y\otimes\sigma_y$ will be selected $\ell\sin^2(2\tau)/4$ times each; and  $\Bid\otimes\sigma_z$ and $\sigma_z\otimes\Bid $ will be chosen $\ell\cos^2(2\tau)/4$ times each.

Every time one of these measurement bases are chosen, we then need to measure each choice $\sigma_k\otimes\sigma_{k'}$ a total of $m_i$ times, where
\begin{equation}
	m_i = \left\lceil  \frac{1}{2\ell\epsilon^2 \cf(k_i,k'_i)^2} \log(2/\delta) \right \rceil.
\end{equation}
For each of the $\Bid\otimes\Bid$ and $\sigma_z\otimes\sigma_z$ measurements, $\lceil 2\log(2/\delta)/\ell\epsilon^2\rceil$ measurements need to be made.
Likewise $\sigma_x\otimes\sigma_x$ and $\sigma_y\otimes\sigma_y$ will require $\lceil 2\sin^2(2\tau)\log(2/\delta)/\ell\epsilon^2\rceil$ measurements; and  $\Bid\otimes\sigma_z$ and $\sigma_z\otimes\Bid $ require $\lceil 2\cos^2(2\tau)\log(2/\delta)/\ell\epsilon^2\rceil$ measurements each.

The steps to then calculate the fidelity for the given states can be found in Following \Refs{FlammiaLiu,daSilva2011,Kliesch2020} and \App{fidelityEstimation}.

\section{Two-qutrit states}\label{twoQutirtProtocols}

Now that we've discussed the existing methods for quantum state verification and direct fidelity estimation, we will consider how these can be extended for larger qudits.
We start with systems of two qutrits here before going on to how these protocols can be generalized for larger bipartite systems in \Sec{GeneralizedSrategies}.
Note that in the case of quantum state verification, we're more interested in generating a strategy that is both efficient and simply scalable, rather than the optimal strategy.

\subsection{Efficient quantum-state verification}

As the dimension of each qudit increases, so does the number of possible substrategies.
Here we will show how an efficient strategy can be generated by just considering two substrategies.
Note also that as the dimension of the qudits increases, the possible target states in the Schmidt decomposition has two degrees of freedom to choose from.
We will therefore decrease this to one degree of freedom and consider the squeezing evolution for two qutrits; doing so allows us to consider and provide results for a subset of interesting and useful states that can be generated in systems of two qutrits.

Considering two qutrits, the time-dependent state from \Eq{GeneralTimeDepState} for $d=3$ is locally equivalent to
\begin{equation}\label{twoQutritState}
\begin{split}
	\ket{\psi(\tau)} = \frac14 & \left(2\cos^2\tau + 6 + \gamma(\tau) \right)^{\frac12}\ket{00} \\
	+& \frac12 |\sin\tau|\ket{11} \\
	 &+ \frac14 \left(2\cos^2\tau + 6 - \gamma(\tau) \right)^{\frac12}\ket{22}, 
\end{split}
\end{equation}
where 
\begin{equation}\label{theGamma}
	\gamma(\tau) = (2 \cos(\tau)+2) \sqrt{\cos(\tau)^2-2 \cos(\tau)+5}.
\end{equation}
It was demonstrated in \Refe{Pallister} that a strategy for a state in the Schmidt decomposition gives equivalent results to the equivalent local rotation of the strategy applied to the actual state.
This is also valid for states of larger dimension.
We can therefore consider the strategy for the state in \Eq{twoQutritState} as the Schmidt decomposition of \Eq{GeneralTimeDepState}.

\begin{figure}[t!]
	\includegraphics[width=\linewidth]{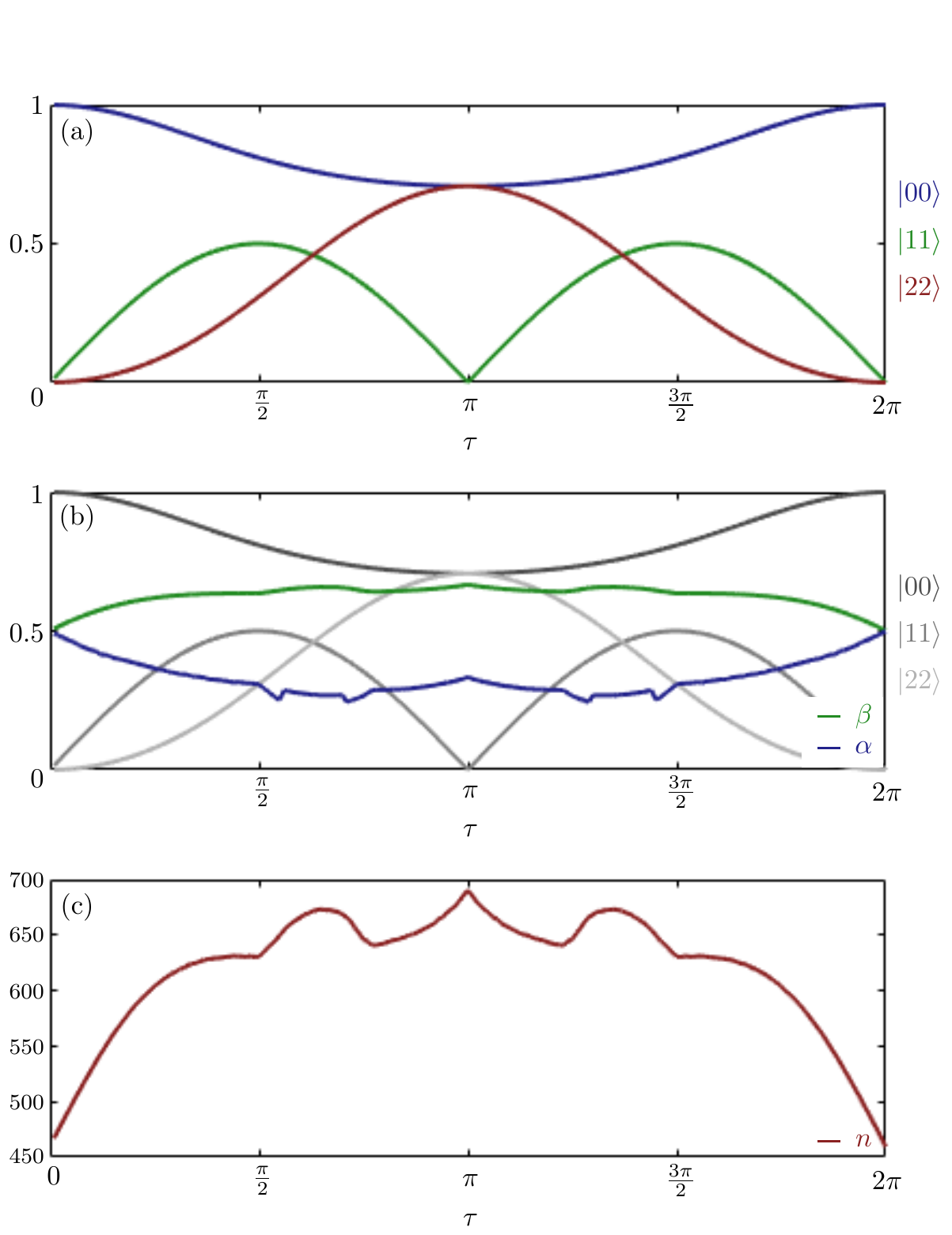}	
	\caption{\label{theState} \Sub{a} shows the elements of the state vector for two-qutrit state through two-qutrit squeezing evolution.
	\Sub{b} shows the values of $\alpha$ from \Eq{TwoQutritStrategyProto} in blue, and the value of $\beta$ from \Eq{twoQutritBeta} in green. 
	These are numerical results where we have optimized the value of $\theta_3$.
	\Sub{c} shows the number of measurements needed to verify two-qutrit state with error rate $\epsilon = 0.01$ and failure probability $\delta = 0.1$.}
\end{figure}

The three parts of the state \Eq{twoQutritState} are shown in \FigSub{theState}{a}, where the blue on the top shows the coefficients in the $\ket{00}$ basis through time.
Likewise the green line underneath shows the coefficients for $\ket{11}$ and the red line is $\ket{22}$.
Like in the two-qubit case, different types of states require more specialized strategies.
At the beginning and end of the evolution the state is separable, and so the optimal strategy is straightforward.
Also like the two-qubit case, at $\tau=\pi$ we yield the state $(\ket{00}+\ket{11})/\sqrt{2}$. 
To verify the state at this point, we simply need to utilize the strategy in \Eq{BellGenStrategy}.
We note that other points in the evolution may themselves require a separate form of a strategy to be more efficient or optimal, however we will now consider a more general form of a two-qutrit strategy.

In order to provide a strategy to verify a general two-qutrit state in this evolution, we need to consider the structure of a generalized strategy.
A strategy for two qutrits can be generated as the convex sum
\begin{equation}\label{GenStrategy}
	\Omega = \sum_{i=1}^8 c_i \Omega_i,
\end{equation}
where is $\Omega_i$ is a rank-$i$ strategy. 
Like for \Eq{GenStrategy2Qubits}, each rank-$i$ strategy, $\Omega_i$, is generated by a weighted sum of projectors $P_i^k$, where each $P_i^k\ket{\psi} = \ket{\psi}$.

We want to construct each $\Omega_i$ so that, given a set of states $\ket{\psi_l^\perp}$ orthogonal to $\ket{\psi}$, we minimize the maximum value of $\bra{\psi_l^\perp}\Omega_i\ket{\psi_l^\perp}$.
After finding each substrategy, we want to find the values of $c_i$ in \Eq{GenStrategy} such that we minimize the maximum value of $\bra{\psi_l^\perp}\Omega\ket{\psi_l^\perp}$.
Now, to simplify the strategy and to create analogy to the two-qubit case, we will consider rank-3 and rank-7 substrategies.

We can then introduce our general strategy 
\begin{equation}\label{TwoQutritStrategyProto}
	\Omega = \alpha \Pz{3} + (1-\alpha) \sum_j \Omega_7^j,
\end{equation}
where rank-3 part is constructed
\begin{equation}
	\Pz{3} = \Proj{00}+\Proj{11}+\Proj{22},
\end{equation}
and the rank-7 strategy is of the form
\begin{equation}\label{genRank7Strat}
	\Omega_7^j = \eta_7^j(\Bid_9 - (\rho_7^j \otimes \sigma_7^j + \rho_7^{j \perp} \otimes \sigma_7^{j \perp} ))
\end{equation}
where $\bra{\psi }\rho_7^j \otimes \sigma_7^j\ket{\psi} = \bra{\psi } \rho_7^{j \perp} \otimes \sigma_7^{j \perp}\ket{\psi} = 0$.

Note that we generalize $\Pz{2}$ for two qubits to $\Pz{3}$ as this provides the same role of always accepting a state that is in the Schmidt decomposition, while rejecting all other orthogonal states.
We then need the second part of the strategy pull the down the probability of accepting one of the orthogonal state that $\Pz{3}$ accepts.
Here we have chosen a rank-7 strategy of the form in \Eq{genRank7Strat}, where more on how we construct part of the strategy can be found in \App{QutritStatElements}.




Following the logic for two qubits, we now consider two general states that are orthogonal to the target state.
To simplify this calculation, let's first consider a general two-qutrit state
\begin{eqnarray}\label{twoQutritTargetState}
	\ket{\psi(\theta_1,\theta_2)} &=& \sin\theta_2\cos\theta_1 \ket{00} \\
	& & + \sin\theta_2\sin\theta_1 \ket{11} + \cos\theta_2 \ket{22},\nonumber
\end{eqnarray}
that is locally equivalent to any two-qutrit pure state.
The state in \Eq{twoQutritState} is generated by setting
\begin{eqnarray}
	\theta_1 &=& \arccos\left(2\left[\frac{\sin^2 \tau}{
	2\cos^2\tau + 6 + \gamma(\tau)} \right]^\frac12\right)\\
	\theta_2 &=& \arctan\left(\frac{1}{4}\left[2\cos^2\tau + 6 - \gamma(\tau)\right]^\frac12\right),
\end{eqnarray}
where $\gamma(\tau)$ is defined in \Eq{theGamma}.
We can then choose two general states that are orthogonal to \Eq{twoQutritTargetState}
\begin{eqnarray}
	\Basis{1} &=& (\cos\theta_1\cos\theta_2\cos\theta_3 - \sin\theta_1\sin\theta_3)\ket{00}\nonumber\\
	& + & (\sin\theta_1\cos\theta_2\cos\theta_3 + \cos\theta_1\sin\theta_3) \ket{11}\nonumber \\
	& - & \sin\theta_2\cos\theta_3 \ket{22}  \label{twoQutritOrth1}\\
	\Basis{2} &=& (\cos\theta_1\cos\theta_2\sin\theta_3 + \sin\theta_1\cos\theta_3)\ket{00}\nonumber\\
	& + & (\sin\theta_1\cos\theta_2\sin\theta_3 - \cos\theta_1\cos\theta_3) \ket{11}\nonumber \\
	& - & \sin\theta_2\sin\theta_3 \ket{22}, \label{twoQutritOrth2} 	
\end{eqnarray}
that hold for any value of $\theta_3$.

To create general separable states that are orthogonal to \Eq{twoQutritState}, we want to create an analog to the state in \Eq{twoQubitStartOrthStates}, this can be done by generating two states that are proportional to \Eq{twoQutritOrth1} and \Eq{twoQutritOrth2} in the relevant entries.
This results in separable states 
\begin{equation}
\begin{split}
	\ket{\phi_1(\boldsymbol{\theta},\boldsymbol{\varphi})} &= \ket{\rho_7(\boldsymbol{\theta},\boldsymbol{\varphi})} \otimes \ket{\sigma_7(\boldsymbol{\theta},\boldsymbol{\varphi})} \\
	\ket{\phi_2(\boldsymbol{\theta},\boldsymbol{\varphi})}&=\ket{\rho_7^\perp(\boldsymbol{\theta},\boldsymbol{\varphi})} \otimes \ket{\sigma_7^\perp(\boldsymbol{\theta},\boldsymbol{\varphi})},
\end{split}
\end{equation}
where
\begin{widetext}
	\begin{eqnarray}
	\ket{\rho_7(\boldsymbol{\theta},\boldsymbol{\varphi})} &=& \frac{1}{\mathcal{N}_\rho^1} \Big( \sqrt{\cos\theta_1\cos\theta_2\cos\theta_3 - \sin\theta_1\sin\theta_3} \ket{0}  +\ue{\ui\varphi_1} \sqrt{\sin\theta_1\cos\theta_2\cos\theta_3 + \cos\theta_1\sin\theta_3} \ket{1}  \nonumber \\
	& &\hspace{7cm} + \;\ue{\ui\varphi_2}\sqrt{-\sin\theta_2\cos\theta_3} \ket{2} \Big)  \\
	\ket{\sigma_7(\boldsymbol{\theta},\boldsymbol{\varphi})} &=& \frac{1}{\mathcal{N}_\rho^1} \Big( \sqrt{\cos\theta_1\cos\theta_2\cos\theta_3 - \sin\theta_1\sin\theta_3} \ket{0} + \ue{-\ui\varphi_1} \sqrt{\sin\theta_1\cos\theta_2\cos\theta_3 + \cos\theta_1\sin\theta_3} \ket{1}  \nonumber \\
	& &\hspace{7cm} + \;\ue{-\ui\varphi_2}\sqrt{-\sin\theta_2\cos\theta_3} \ket{2} \Big) \\
	\ket{\rho_7^{\perp}(\boldsymbol{\theta},\boldsymbol{\varphi})} &=& \frac{1}{\mathcal{N}_\rho^2} \Big( \sqrt{\cos\theta_1\cos\theta_2\sin\theta_3 + \sin\theta_1\cos\theta_3} \ket{0} + \ue{\ui\varphi_1} \sqrt{\sin\theta_1\cos\theta_2\sin\theta_3 - \cos\theta_1\cos\theta_3} \ket{1}  \nonumber\\
	& &\hspace{7cm} + \;\ue{\ui\varphi_2}\sqrt{-\sin\theta_2\sin\theta_3} \ket{2} \Big) \\
	\ket{\sigma_7^{\perp}(\boldsymbol{\theta},\boldsymbol{\varphi})} &=& \frac{1}{\mathcal{N}_\rho^2} \Big( \sqrt{\cos\theta_1\cos\theta_2\sin\theta_3 + \sin\theta_1\cos\theta_3} \ket{0}  + \ue{-\ui\varphi_1} \sqrt{\sin\theta_1\cos\theta_2\sin\theta_3 - \cos\theta_1\cos\theta_3} \ket{1}  \nonumber \\
	& &\hspace{7cm} + \;\ue{-\ui\varphi_2}\sqrt{-\sin\theta_2\sin\theta_3} \ket{2} \Big),
\end{eqnarray}
\end{widetext}
whose construction is shown in \App{QutritStatElements}, and $\mathcal{N}_\rho^1$ and $\mathcal{N}_\rho^2$ are the appropriate normalizations of these states.
\Eq{genRank7Strat} then becomes
\begin{equation}
	\Omega_7^j = \eta_7^j\left(\Bid_9 - \sum_n\Proj{\phi_n(\boldsymbol{\theta},\boldsymbol{\varphi})} \right)
\end{equation}

We now have the two parts of the strategy in \Eq{TwoQutritStrategyProto}.
To calculate the sum of the rank-7 strategies, we need to average over all values of $\varphi_1$ and $\varphi_2$.
Like before for the phase terms, we can take $ \varphi_1, \, \varphi_2 = \{0, 2\pi/3, 4\pi/3 \}$. 
Averaging over these discrete values is equivalent to integrating over both $\varphi$ degrees of freedom.

Next we need to consider what values of $\theta_3$ we need for each $\tau$.
Here we will simply choose a fixed value of $\theta_3$, rather than averaging over different values like we did with the $\varphi$ degrees of freedom. 
Our goal is to devise an efficient strategy that is simply scalable to higher-dimensional systems, and the procedure is vastly simplified by choosing a fixed value of $\theta_3$. 

Note that when generalizing to qudits in \Sec{gen2Qudits}, we will simply let the extra $\theta$ degrees of freedom be 0; in this section however, we will choose the optimal value for a single choice of $\theta_3$ for each time step -- note that this does not necessarily mean that the overall strategy is optimal.

Given a value of $\theta_3$, we can now build an orthonormal basis to describe the construction of our strategy.
These are the target state and 8 orthogonal states $\Basis{i}$ for $1\leq i \leq 8$, where $\Basis{1}$ and $\Basis{2}$ are in  \Eq{twoQutritOrth1} and \Eq{twoQutritOrth2} respectively.
The other six states are $\Basis{3}=\ket{01}$, $\Basis{4}=\ket{02}$, $\Basis{5}=\ket{10}$, $\Basis{6}=\ket{12}$, $\Basis{7}=\ket{20}$, $\Basis{8}=\ket{21}$.

Considering the strategy in the form of \Eq{TwoQutritStrategyProto}, we now need to use these states to find an efficient strategy.
Like in the two qubit-case, the $\Pz{3}$ term accepts the $\ket\psi$, $\Basis{1}$ and $\Basis{2}$ with certainty and rejects all the other states that are orthogonal to these three.
$\Omega_7$ accepts $\ket\psi$ with certainty and accepts all other states with a non-zero probability.
It provides a role of pulling down the probability of accepting  $\Basis{1}$ and $\Basis{2}$, while increasing the probability of accepting the other orthogonal states.
All that is required is to optimize $\alpha$ to ensure the lowest chance of accepting a state orthogonal to $\ket\psi$.

We now apply this strategy to the target state.
At the beginning of the evolution, $0<\tau<\pi/2$, setting $\theta_3=0$ is the optimal choice of $\theta_3$.
Given this choice of $\theta_3$, we will now consider the how the components of the strategy interact with the basis states.
For the $\Omega_7$ part of the strategy, the largest value of $\bra{\basisPsi{i}}\Omega_7\ket{\basisPsi{i}}$ is for $\Basis{8}$.
Likewise, the largest value for $\bra{\basisPsi{i}}\Pz{3}\ket{\basisPsi{i}}$ is for $\Basis{2}$.
To optimize the strategy, we then need to solve $\bra{\basisPsi{8}}\Omega\ket{\basisPsi{8}} = \bra{\basisPsi{2}}\Omega\ket{\basisPsi{2}}$.
The resulting solution of $\alpha$ is shown in the blue line in \FigSub{theState}{b}.
For the rest of the evolution, we then numerically optimize over values of $\theta_3$.

After calculating the optimal value of $\alpha$ for the given construction, we can then calculate the corresponding value of $\beta$, were
\begin{equation}\label{twoQutritBeta}
	\beta = \max_i  \bra{\basisPsi{i}}\Omega\ket{\basisPsi{i}},
\end{equation}
is the highest probability that an orthogonal state will be accepted.
We have shown the values of $\beta$ over time in \FigSub{theState}{b} in green.

Taking the values of $\beta$, we can then calculate the optimal number of measurements needed to verify that a given state has been generated, up to some error $\epsilon$ and some failure probability $\delta$.
Given these parameters, we need to take
\begin{equation}\label{nOpt}
	n \geq \frac{\ln\delta^{-1}}{ \ln \left[ \frac{1}{1-\epsilon(1-\beta)} \right] } 
\end{equation}
measurements~\cite{Pallister,Kliesch2020,ZhuHayashi2019-2}.
Saturating this inequality, we get the values of $n$ in \FigSub{theState}{c}, where we have chosen $\epsilon=0.01$, $\delta=0.1$.

The value of $\beta$ at each point is important to understand and to minimize, we can see in \Eq{nOpt} its role in lowering the number of measurements needed to verify a target state.
We can also see this by considering a given pure state 
\begin{equation}
	\ket{\Psi} = \sqrt{1-\epsilon} \ket{\psi} + \sqrt{\epsilon} \ket{\psi^\perp},
\end{equation}
where $\ket{\psi}$ is the target states and $\ket{\psi^\perp}$ is some state orthogonal to the target state -- that could be any superposition of $\ket{\psi^\perp_i}$ discussed earlier.

Applying this erroneous state to the strategy we yield 
\begin{equation}
\begin{split}
	\bra{\Psi}\Omega\ket{\Psi} &= 1-\epsilon + \epsilon \bra{\psi^\perp}\Omega\ket{\psi^\perp} \\
	&= 1- \epsilon(1-\beta)
\end{split}
\end{equation}
which can be seen by noting the requirement $\Omega\ket{\psi}= \ket{\psi}$.
If $\beta=1$, then this equals 1, resulting in a state that is demonstrably not the target state being accepted by the strategy.
Since $\beta=1$ nowhere, this strategy can successfully distinguish the target state from any other state.

Note that \FigSub{theState}{b} and~\Sub{c} only provide the answers when using the strategy in \Eq{TwoQutritStrategyProto}.
As we have already shown, certain states are special cases and can be verified more efficiently using a modified strategy.
This has already been shown for $\tau = 0,\,\pi,\,2\pi$ in \Sec{specialCases}.
Explicitly, keeping the same values for $\epsilon$ and $\delta$, this reduces the number of measurements needed from $460$ to $229$ for $\tau = 0,\,2\pi$; and from $695$ to $345$ for $\tau=\pi$.



\subsection{Fidelity estimation}

\begin{figure}
	\includegraphics[width = \linewidth]{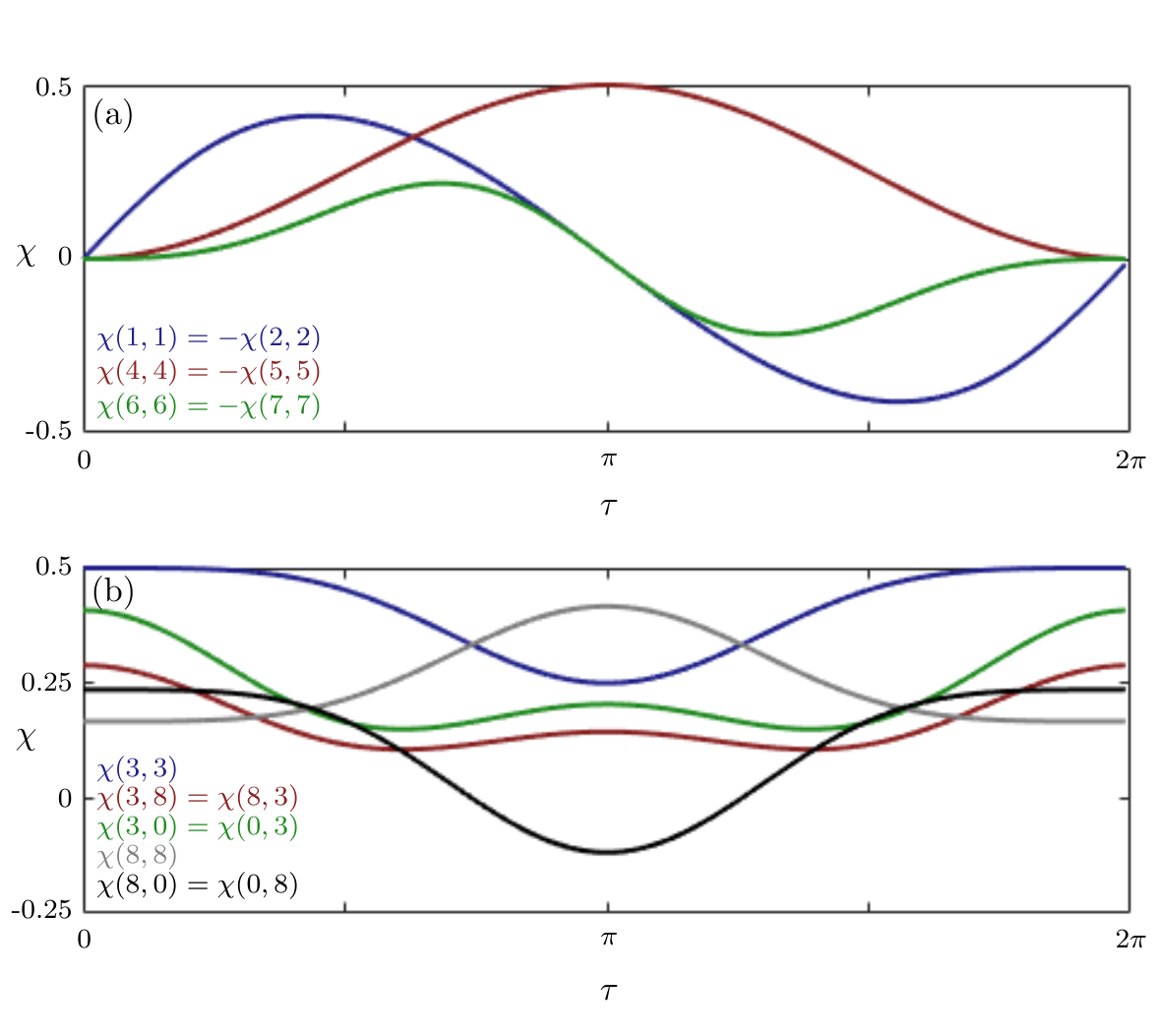}
	\caption{\label{theTwoQutritCharacteristicFunction}
	Elements of the two-qutrit characteristic function of the state \Eq{twoQutritState} throughout the two-qutrit squeezing evolution.
	\Sub{a} shows six of the elements, which are constructed from the off-diagonal generators of \su{3}.
	These come in pairs, half of the elements and -1 times the other half.
	\Sub{b} shows the elements of the characteristic function that are generated from the diagonal operators.
	Note that $\chi(0,0) = 1/3$ always and so is not included in these plots.}
\end{figure}
 
\begin{figure}
	\includegraphics[width=\linewidth]{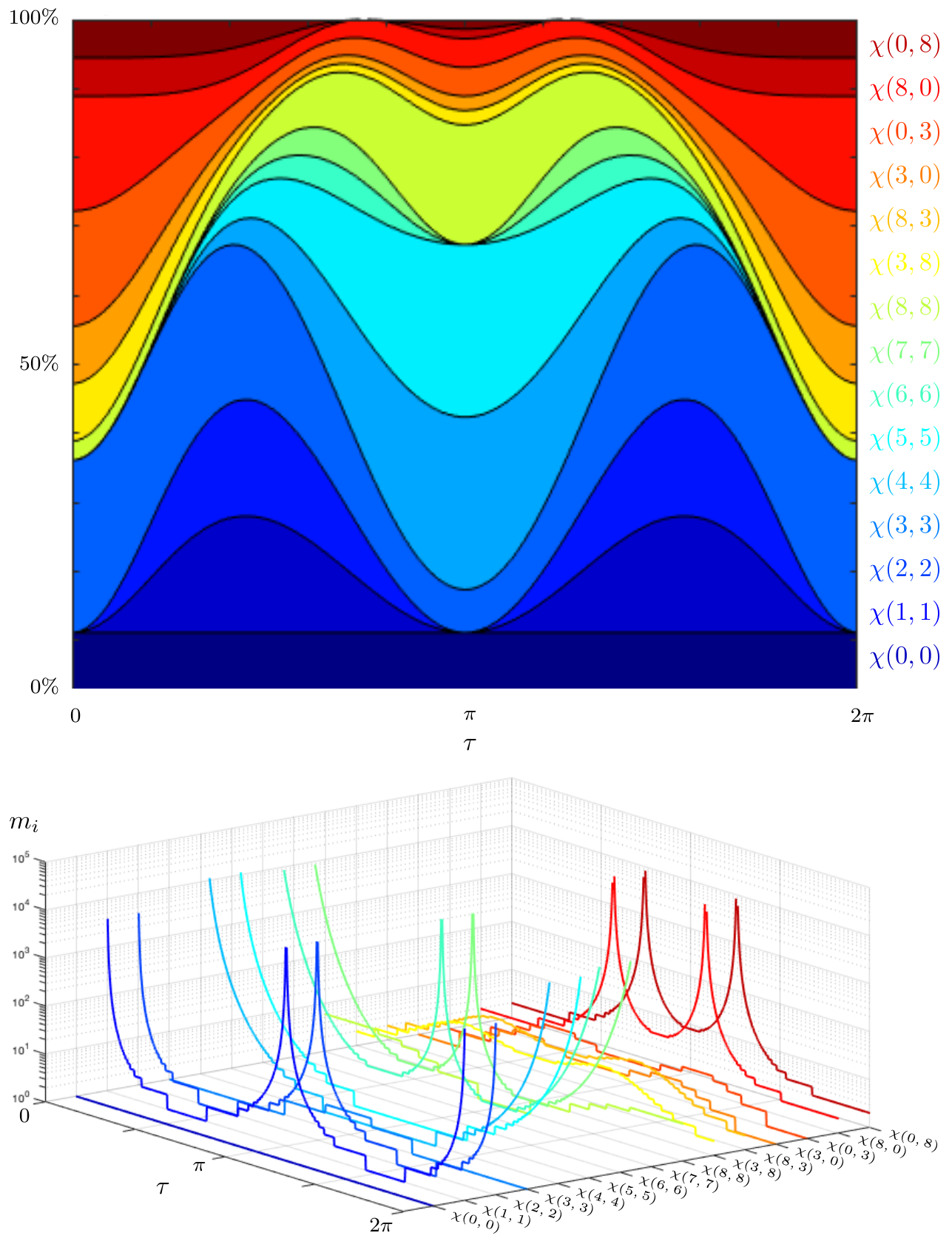}
	\caption{\label{twoQutritFidelityMeasurement} Direct fidelity estimation of two-qutrit states during the squeezing evolution.
	\Sub{a} shows the average proportion each measurement basis is chosen, where each color represents a different value of the characteristic function, where the values representing the colors are shown on the right-hand side.
	These measurements are shown for each point in the evolution, that goes along the horizontal axis.
	The proportion each basis takes along the vertical axis then corresponds to the probability any given basis has of begin chosen for measurement.
	\Sub{b} shows the number of measurements needed when each of the bases are chosen.
	Note that the vertical axis is on a logarithmic scale.}
\end{figure}

The fidelity estimation protocol for two qubits can easily be extended for a two-qutrit system.
The general protocol to do this can be found in \App{fidelityEstimation}. 
We give now given an explicit example of two-qutrit systems here.

First we need to begin with a characteristic function.
This first requires a basis that is a generalization of the Pauli operators for qubits.
The generalization of the Pauli operators can be done in various ways, where it is common to either choose the Weyl algebra or the \su{N} algebra.
Here we will consider the \su{3} algebra, also known as the Gell Mann matrices.
The \su{3} algebra is commonly represented by the matrices
\begin{equation}
\begin{split}
	\Lambda_3^1 &= \begin{pmatrix} 0 & 1 & 0 \\ 1 & 0 & 0 \\ 0 & 0 & 0\end{pmatrix}, \; \Lambda_3^2 = \begin{pmatrix} 0 & -\ui & 0 \\ \ui & 0 & 0 \\ 0 & 0 & 0\end{pmatrix} \\
	\Lambda_3^4 &= \begin{pmatrix} 0 & 0 & 1 \\ 0 & 0 & 0 \\ 1 & 0 & 0\end{pmatrix}, \; \Lambda_3^5 = \begin{pmatrix} 0 & 0 & -\ui \\ 0 & 0 & 0 \\ \ui & 0 & 0\end{pmatrix}\\
	\Lambda_3^6 &= \begin{pmatrix} 0 & 0 & 0 \\ 0 & 0 & 1 \\ 0 & 1 & 0\end{pmatrix}, \; \Lambda_3^7 = \begin{pmatrix} 0 & 0 & 0 \\ 0 & 0 & -\ui \\ 0 & \ui    & 0\end{pmatrix}\\
	 \Lambda_3^3 &= \begin{pmatrix} 1 & 0 & 0 \\ 0 & -1 & 0 \\ 0 & 0 & 0\end{pmatrix},\;\Lambda_3^8 = \frac{1}{\sqrt{3}}\begin{pmatrix} 1 & 0 & 0 \\ 0 & 1 & 0 \\ 0 & 0 & -2\end{pmatrix}.
\end{split}
\end{equation}

Along with the identity operator $\Lambda_3^0 = \Bid$ we have an informationally complete Weyl function for two qutrits
\begin{equation}
	\chi_\rho(k,k') = \frac{1}{\mathsf{N}_{k}\mathsf{N}_{k'}}\Trace{\rho\,\Lambda_3^k\otimes \Lambda_3^{k'}}
\end{equation}
where 
\begin{equation}
	\mathsf{N}_{k} = \begin{cases}
		\sqrt{3} & \text{for } k = 0,\\
		\sqrt{2} & \text{otherwise.}
	\end{cases}
\end{equation}

For our target state from \Eq{twoQutritState}, at most 15 element of the characteristic function are non-zero.
Where at $\tau=0,\pi,2\pi$ even fewer are non-zero.
The elements of the characteristic function over time are shown in \Fig{theTwoQutritCharacteristicFunction}.

In \Fig{theTwoQutritCharacteristicFunction} we separate the elements of the characteristic function into the diagonal and off-diagonal operators.
This is because all of the diagonal operators needed are symmetric, \textit{i.~e.~} $k=k'$.
This then generates six of the elements of the characteristic function.
These can be found in \FigSub{theTwoQutritCharacteristicFunction}{a}, where only three curves are shown since $\chi(2,2) = -\chi(1,1)$, $\chi(5,5) = -\chi(4,4)$ and $\chi(7,7) = -\chi(6,6)$

The other nine elements of the characteristic function come from all combinations of the three diagonal operators.
This produces five the unique values shown in \FigSub{theTwoQutritCharacteristicFunction}{b} where $\chi(i,j) = \chi(j,i)$ for $i,j\in\{0,3,8\}$.
Note that $\chi(0,0)=1/3$ always, and so isn't included in \FigSub{theTwoQutritCharacteristicFunction}{b}.

Now that we have the values for the characteristic function of the target state at each point in the evolution, we can now consider how many measurements are needed for fidelity estimation.
Given $\ell$ choices of basis in \Eq{theEll}, we can calculate the probability each basis will be chosen.
This is shown visually in \FigSub{twoQutritFidelityMeasurement}{a}, where we have presented the measurement bases as the average percentages for each time step.
Each measurement basis is represented by a unique color in the contour plot.
The area each color takes up in the vertical axis is the average proportion of the $\ell$ measurements each basis will be chosen, for each moment in the time of the evolution.

Because $\chi(0,0) = 1/3$ always, we can see how measurement of $\Bid\otimes\Bid$ has an $11\%$ chance of being chosen.
Also note that at certain points, far fewer bases need to be taken, where at $\tau = 0,\pi,2\pi$, there are four fewer measurement bases that need to be considered.

Now we need to consider how many measurements are needed when each basis is chosen.
This is calculated by
\begin{equation}
	m_i = \left\lceil  \frac{2}{\mathsf{N}_{k}^2\mathsf{N}_{k'}^2\ell\epsilon^2 \cf(k_i,k'_i)^2} \log(2/\delta) \right \rceil.
\end{equation}
Choosing $\epsilon = 0.01$ and $\delta=0.1$ to calculate $m_i$ for each point in the evolution, we yield  the results in \FigSub{twoQutritFidelityMeasurement}{b}. 
Again each basis is represented by a unique color, where the vertical axis shows the number of measurements needed on a log axis. 
When value of the characteristic function is particularly small, the number of measurements needed increase with $1/\chi^2$, resulting in some large peaks in \FigSub{twoQutritFidelityMeasurement}{b}.

This gives the number of measurements needed to calculate the fidelity, by following the procedure laid out in \App{fidelityEstimation} one can then estimate the fidelity of any of the states produced in this evolution.

\section{Two-qudit quantum state verificaiton}\label{GeneralizedSrategies}\label{gen2Qudits}
Given the previous results for two-qubit systems and the extension to two-qutrit systems, we now further generalize these results to two qudits of arbitrary dimension.
We will focus on the construction of an efficient procedure to verify a general two-qudit state, where the generalization of the fidelity estimation protocol is given in \App{fidelityEstimation}.


To generalize the construction of this strategy to a bipartite system of any dimension, we first need to define the substrategies.
Like in \Eq{TwoQutritStrategyProto}, the strategy is built from two parts. 
The first part is the generalization of the $\Pz{d}$ operator, this is simply
\begin{equation}
	\Pz{d} = \sum_{i=0}^{d-1} \Proj{dd}.
\end{equation}

The second part is to generalize the rank-7 part of the strategy in \Eq{TwoQutritStrategyProto}.
In general we will require a rank-$r$ strategy, where $r = (d^2-d+1)$. 
This rank-$r$ strategy needs to itself be constructed from separable states that are orthogonal to the target state. 
We then in effect build a rank-$d-1$ strategy from the projectors of these states, that we then subtract from the identity operator, generating a rank-$r$ strategy.

In \Eq{twoQutritOrth1} and \Eq{twoQutritOrth2} we gave two general states that are orthogonal to the target state.
We then went onto making separable versions of these states. 
Specifically, this involved creating a separable state, $\ket{\phi_i^d(\boldsymbol{\theta},\boldsymbol{\varphi})}$, where
\begin{equation}
	\langle kk \ket{\phi_i^d(\boldsymbol{\theta},\boldsymbol{\varphi})} =\frac{\langle kk \Basis{i}}{\sqrt{\IP{\phi_i^d(\boldsymbol{\theta},\boldsymbol{\varphi})}{\phi_i^d(\boldsymbol{\theta},\boldsymbol{\varphi})}}},
\end{equation}
\emph{i.~e.} the $\ket{kk}$ elements of the separable states are proportional to $\langle kk \Basis{i}$.

To generalize these states, we first need a reliable method to create an orthonormal basis for a $d$-dimensional system.
One way is to adapt the coherent-state approach for states with an \SU{d} symmetry~\cite{Nemoto2000,TilmaKae1}.
The procedure to do this is given in \App{EulerAngles}.

The goal is to use the Euler-angle construction to compose a coherent state whose elements are the Schmidt coefficients of the target state.
That is, given the Schmidt decomposition of a general two-qutrit target state
\begin{equation}\label{twoQutritSchmidt}
	\ket{\psi} = \sum_{k=1}^d c_k\ket{kk},
\end{equation}
we wish to create a coherent state
\begin{equation}\label{singleQutritCoherent}
	\ket{\boldsymbol{\theta}} = \sum_{k=1}^d c_k\ket{k},
\end{equation}
finding the solution for the $\boldsymbol{\theta}$ degrees of freedom.
In the spin-coherent state construction, this is defined as the Euler operator applied to the highest-weighted state, $\ket{d-1}$.

\begin{figure}[t!]
\includegraphics[width=\linewidth]{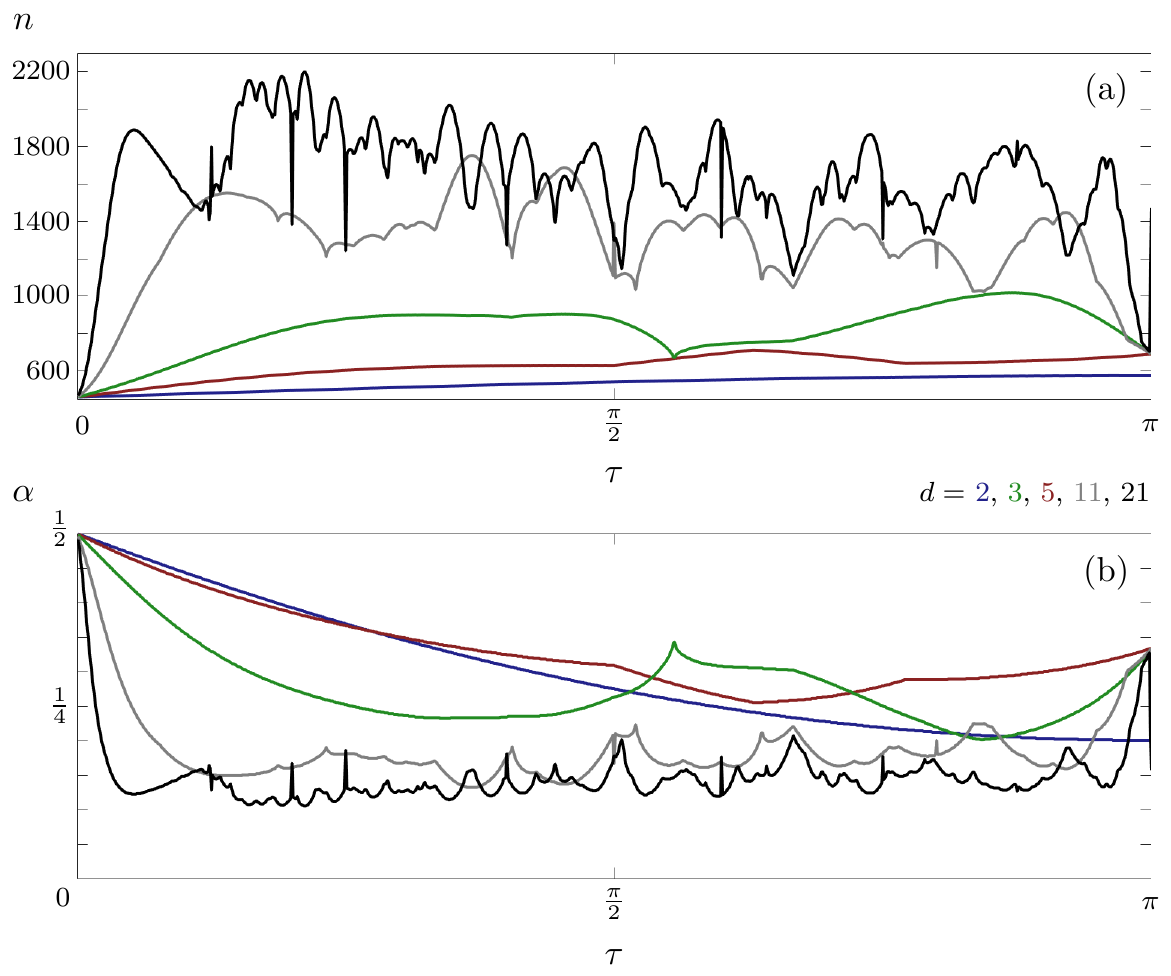}
\caption{\label{nOptqOptGeneral} The number of measurements needed to verify a state with $\epsilon = 0.99$ and $\delta = 0.1$.
\Sub{a} shows the number of measurements needed for the two-qudit squeezing evolution for $d = 2,\,3,\,5,\,11,$ and $21$ for $0\leq\tau\leq\pi$.
\Sub{b} is the value of $\alpha$ for each case, that shows the balance between the $\Pz{d}$ and $\Omega_r$ substrategies.
Note that we have only shown half the evolution here as the second half identical, only reversed.}
\end{figure}

We can then generate an orthonormal basis by applying the Euler operator to the other computational basis states, $\ket{k}$ where $0\leq k < d-1$.
By reversing the step from \Eq{singleQutritCoherent} to \Eq{twoQutritSchmidt} we can convert these single-qudit orthogonal states into two-qudit states in the Schmidt decomposition, which are the generalization of $\ket{\psi^\perp_1}$ and $\ket{\psi^\perp_2}$ from \Eq{twoQutritOrth1} and \Eq{twoQutritOrth2}.
This along with the states $\ket{k_1 k_2}$, where $k_1 \neq k_2$, builds a basis for a two-qudit strategy.

From the first $d-1$ orthogonal states $\ket{\psi^\perp_i}$, and by following the procedure in \App{EulerAngles}, we can then get a general set of separable states  $\ket{\phi_i^d(\boldsymbol{\theta},\boldsymbol{\varphi})} $ for $i=1...d-1$, that are orthogonal to the target state.

The steps are now similar to the two-qutrit case, where we now generate
\begin{equation}
	\Omega_{r} =   \frac{1}{3^{d-1}}\sum_{\boldsymbol{\varphi}}\left(\Bid_d - \sum_{n=1}^{d-1}\Proj{\phi_n^d(\boldsymbol{\theta},\boldsymbol{\varphi})} \right).
\end{equation}
The sum over $\boldsymbol{\varphi}$ is an average over the phases, where we restrict this to each $\varphi_i = \{0, 2\pi/3, 4\pi/3 \}$.
This results in $3^{d-1}$ different phases.

We now need to build the full strategy
\begin{equation}
	\Omega = \alpha \Pz{d} + (1-\alpha) \Omega_{r},
\end{equation}
where we want to optimize over $\alpha$ and $\overline{\boldsymbol{\theta}}$, where $\overline{\boldsymbol{\theta}} =  \{ \theta_{d-1}, ..., \theta_{d(d-1)/2}\}$.
Although a more efficient strategy can be found by optimizing over $\overline{\boldsymbol{\theta}}$, it is a far more simple construction to just set $\overline{\boldsymbol{\theta}} = 0$.
Given this compromise in optimality, we have calculated the optimal values for $\alpha$ in these cases numerically.

We have applied this to the two-qudit squeezed Hamiltonian for $d=2,3,5,11$ and $21$, showing the results for the number of measurements needed and the values of $\alpha$ in \Fig{nOptqOptGeneral}.
Note that we have considered $0\leq\tau\leq\pi$ here, as the results are just reversed for the second half of the evolution.
From the value of $\alpha$ we can then construct an efficient strategy.
From this strategy, we just need to find the second largest eigenvalue, $\beta$.
Applying this value of $\beta$ to \Eq{nOpt} then yields the number of measurements needed to verify that the desired state is generated within some choice of $\epsilon$ and $\delta$.
\FigSub{nOptqOptGeneral}{b} shows this calculated for $\epsilon=0.01$ and $\delta=0.1$.

\section{Conclusions}
Here we have provided procedures to generate efficient strategies to verify -- and to directly estimate the fidelity of -- two-qudit states, both with local measurements.
Although not optimal, this quantum-state verification procedure is simply extendable to a bipartite system of any size.

The strategies provided can be used to verify any two-qudit state, where a two-qudit target state in the Schmidt decomposition has $d-1$ degrees of freedom.
We gave examples of how these strategies perform in the two-qudit squeezing evolution.
The states produced in this evolution are of particular interest as both two-qudit squeezed states and various entangled states are generated.
States that are of particular interest in a quantum technology setting.

We also note that the techniques used here can be used to verify a single-axis twisting Hamiltonian on a single qubit, by utilizing the Choi-Jamiołkowski isomorphism and verifying the two-qutrit Choi matrix~\cite{YeChaoLiu2020}.
Both the single-qudit and two-qudit squeezing evolutions have themselves seen significant interest, whether in creating large entangled state~\cite{Song2019} or when entangling Bose Einstein condensates~\cite{Byrnes2012,Byrnes2013,MengXin2021}.
It is therefore conceivable that we could use the strategies provided in this work to experimentally verify the creation of these bipartite squeezed and entangled Bose Einstein condensates.

We have focussed here on providing a simple procedure to generate a strategy for two qudit states, but this may have sacrificed optimality.
Future work is needed to optimize these strategies, which will require optimizing over the degrees of freedom; whether by averaging over many values or by finding an optimal value.
Further optimization may also come in the form of introducing more and different types of components to the strategy.

We also note that an adaptation of the strategy could be useful when applied to a bipartite infinite-dimensional system, if we choose to truncate the Hilbert space to some degree.
We could then apply this method to verify the creation of two-mode squeezed states.
However, this would depend on the possibility of measuring the necessary observables in such systems.

\acknowledgments
We would like to thank Noah Linden for general support and enlightening conversation.
We would also like to thank Martin Kliesch for interesting discussions.
The author acknowledges support from the EPSRC grant number EP/T001062/1 (EPSRC Hub in Quantum Computing and Simulation).

\bibliography{refs.bib}

\clearpage
\appendix

\section{Fidelity estimation}\label{FidelityEstimation}\label{fidelityEstimation}

Here we will discuss how the direct fidelity estimation protocols in \Refs{FlammiaLiu,daSilva2011} can be extended to higher-dimensional discrete systems.
A similar result has been presented in terms of a frame theoretic construction~\cite{KlieschLectureNotes}.
By following these results and using an appropriate choice of operators for a characteristic function, this is a fairly simple extension. 
Further, the fidelity estimation of a bipartite states is particularly straightforward when considering the Schmidt decomposition of the state.

We begin by introducing the bipartite characteristic function for two qubits
\begin{equation}\label{TwoQubitCharacteristic}
	\cf_\rho(k,k') = \frac{1}{2}\Trace{\rho\;\sigma_k\otimes\sigma_{k'}}.
\end{equation}
To generalize this to two-qudit states, we first need to generalize the characteristic function to higher-dimensional states.
To build a discrete characteristic function we need a set of $d^2$ operators, we denote by $\mathcal{D}_d(k)$, that fully describe a single-qudit state.
Such a choice is not unique, as there are many generalizations to the Pauli operators to higher-dimensional systems. 

We show two of the most common approaches, the Weyl algebra and the $\su{d}$ algebra (or the generalized Gell Mann matrices).
The construction of which can be found in \App{App:CharacteristicFunctions}.

We note here that it is important that we normalize the characteristic function such that
\begin{equation}\label{characteristicFunctionNormalization}
	\sum_k |\cf_\rho(k)|^2 = 1,
\end{equation} 
which is different from more standard constructions of a characteristic function, found in texts such as~\Refs{Gross2016,VOURDAS1997367,Rundle2018,Rundle2021}.

Given a suitable choice of representation to build a characteristic function, \Eq{TwoQubitCharacteristic} can be generalized to
\begin{equation}
	\cf_\rho(\mathbf{k}) = \frac{1}{\mathcal{N}(\mathbf{k})}\Trace{\rho\;\mathcal{D}_d(\mathbf{k})},
\end{equation} 
where $\mathbf{k} = (k,k')$, $\mathcal{D}_d(\mathbf{k}) = \mathcal{D}_d(k)\otimes\mathcal{D}_d(k')$ and $\mathcal{N}(\mathbf{k}) = \mathcal{N}(k)\mathcal{N}(k')$ is the normalization to ensure \Eq{characteristicFunctionNormalization} is satisfied. 
This normalization can be simply calculated by requiring $\Trace{\mathcal{D}_d(k)\mathcal{D}_d(k)^\dagger}/\mathcal{N}_d(k)^2=1$.
In practice this results in $\mathcal{N}(k) = \sqrt{d}$ in the Weyl algebra case, and  $\mathcal{N}(k)  = \sqrt{\Trace{\mathcal{D}_d(k)^2}}$ in the \su{d} case.
Note that we also require that 
$\Trace{\mathcal{D}_d(k_1)\mathcal{D}_d(k_2)^\dagger}=0$ when $k_1 \neq k_2$.

From this, one can then calculate the fidelity
\begin{equation}
	F(\rho_1,\rho_2) = \sum_{\mathbf{k}} \chi_{\rho_1}(\mathbf{k})\,\overline{\chi}_{\rho_2}(\mathbf{k})
\end{equation}
where we note that the characteristic function can be complex-valued, which is generally the case when using the Weyl algebra.
Now that the characteristic function has been defined, we will now follow the protocol laid out in \Refe{FlammiaLiu} for direct fidelity estimation. 

Begin by choosing $\ell = \lceil 1/\epsilon^2\delta \rceil$ values of $\mathbf{k}$, where $\epsilon$ is the adaptive error and $\delta$ is the failure probability.
Each sample $\mathbf{k}_i$ will be chosen with the probability $|\cf(\mathbf{k}_i)|^2$, for $\mathbf{k}_1, ..., \mathbf{k}_\ell$.
We then need to measure $\mathcal{D}_d(\mathbf{k}_i)$ a total of $m_i$ times, where
\begin{equation}
	m_i = \left\lceil  \frac{2}{\ell\epsilon^2\mathcal{N}(\mathbf{k}_i)^2\,\cf(\mathbf{k}_i)^2} \right \rceil\log(2/\delta).
\end{equation}
Note that different approaches need to be taken here, depending on your choice of operators.

For each $i$, we then define the random variable
\begin{equation}
	X(\mathbf{k}_i) = \frac{\chi_{\rho_2}(\mathbf{k}_i)}{\chi_{\rho_1}(\mathbf{k}_i)} = \frac{\Trace{\rho_2\,\mathcal{D}_d(\mathbf{k}_i)}}{\mathcal{N}(\mathbf{k}_i)\chi_{\rho_1}(\mathbf{k}_i)}
\end{equation}
such that $E[X(\mathbf{k}_i)] = \chi_{\rho_1}(\mathbf{k}_i)\,\overline{\chi}_{\rho_2}(\mathbf{k}_i)$.
We can then calculate the estimator for $X(\mathbf{k}_i)$ by
\begin{equation}\label{XEstimatorFidelity}
	\tilde{X}(\mathbf{k}_i) =  \frac{1}{m_i \mathcal{N}(\mathbf{k}_i)\,\cf(\mathbf{k}_i)}\sum_{j=1}^{m_i}M_{ij}.
\end{equation}

Repeating this process, we can then calculate the estimator for the fidelity
\begin{equation}\label{YEstimatorFidelity}
	\tilde{Y} = \frac{1}{\ell} \sum_{i=1}^\ell \tilde{X}(\mathbf{k}_i).
\end{equation}
This yields an estimator to the fidelity within the range $[\tilde{Y}-2\epsilon,\tilde{Y}+2\epsilon]$ with probability greater than $1 -2\delta$.
More details on how to construct these for specific choice can be found in \App{App:CharacteristicFunctions}.

Because the measurement process involves choosing $\mathbf{k}$ with probability $P(\mathbf{k})=\cf_{\rho_1}(\mathbf{k})^2$, any points where $\cf_{\rho_2}(\mathbf{k})=0$ requires no measurement to be taken.
Given the structure of the Schmidt decomposition, this results in a big reduction in the number of measurement bases.
If all the measurements are needed, we would need the full $d^4$ described by the representation, however this is reduced to at most $d^3$ needing to be taken if considering the Weyl operators, and $2d^2-d$ if considering the \su{2} algebra.

\section{Bases to measure the characteristic function}\label{App:CharacteristicFunctions}
The generalization of the characteristic function in \Sec{FidelityEstimation} to higher-dimensional systems can take multiple forms.
Here we consider two approaches to construct the characteristic function with discrete degrees of freedom.
These will be the Weyl algebra, also known as the clock and shift matrices.
And the generators for \su{d}, also known as the generalized Gell-Mann matrices.

Note that one could also describe the characteristic function of a finite-dimensional system with continuous degrees of freedom, see for example \Refs{Rundle2018,Rundle2021}; one could then use a procedure similar to the continuous-variable calculation of the fidelity in \Refe{daSilva2011} to perform direct fidelity estimation.
Such an approach may be beneficial for particularly large Hilbert spaces, but is beyond the scope of this paper.

\subsection{\su{d} algebra}\label{sudAlgebra}

We will now focus on the construction of the characteristic function through the \su{d} generalized Gell-Mann algebra.
The diagonal generators are constructed
\begin{equation}
	\lambda_{n^2-1}^d =\left( \sqrt{\tfrac{2}{n(n-1)}}\sum_{i=1}^{n-1} \Proj{i}\right) - \sqrt{\tfrac{2(n-1)}{n}}\Proj{n},
\end{equation}
for $n \geq 2$, and $\lambda_{0}^d=\Bid$.

The off-diagonal generators are simply the \su{2} generators, spanning the full \su{d} space.
We can separate the off-diagonal operators into two types, the generalization of $\Sx$ and $\Sy$, where
\begin{equation}
\begin{split}
	\mathsf{X}^d_{i,j} &= \ket i \bra{j-1} + \ket{j-1} \bra i \\
	\mathsf{Y}^d_{i,j} &= \ui \ket i \bra{j-1} - \ui \ket{j-1} \bra i 
\end{split}
\end{equation}
where $1\leq i \leq d-1$,$1 \leq j \leq i$.
Formally these operators are then ordered with respect to an \su{2} subspace of \su{d}, see \Refs{TilmaKae1,Rundle2018} for more detail.
However the ordering is not that important as long as it is kept track of.
We will then assume some order and refer to the algebra by the matrices $\lambda^d_k$.

The characteristic function is then
\begin{equation}
	\chi_\rho(k) = \frac{1}{\mathcal{N}(k)}\Trace{\rho \, \lambda^d_k},
\end{equation}
where to satisfy \Eq{characteristicFunctionNormalization} the normalization is
\begin{equation}
	\mathcal{N}(k) = \begin{cases} \frac{1}{\sqrt{d}},  & \text{for } k=0,\\
 \frac{1}{\sqrt{2}}, & \text{for } 1\leq k \leq d^2-1. \end{cases}
\end{equation}

The characteristic function is informationally complete, and the full state can be reconstructed by
\begin{equation}
	\rho = \sum_k \frac{1}{\mathcal{N}(k)} \chi_\rho(k)\lambda^d_k,
\end{equation}
where $\mathcal{N}(k)$ is the normalization required by the operators used to define the Weyl function. 

The fidelity of some state, $\sigma$, with respect to a target state, $\rho$, can then be calculated through the characteristic functions, where
\begin{equation}
	\Trace{\rho\sigma} = \sum_k \chi_\rho(k) \chi_\sigma(k).
\end{equation}

From this we can then follow the procedure in \Sec{FidelityEstimation}.
It is now worth following the proof of this procedure as outlined in \Refe{FlammiaLiu} and point how and why this still holds for the \su{d} algebra.
We want to prove that the fidelity is estimated with in the range  $[\tilde{Y}-2\epsilon,\tilde{Y}+2\epsilon]$ with probability greater than $1 -2\delta$.
Equivalently we can express this as 
\begin{equation}
	\mathrm{Pr}\left[|\tilde{Y} - F(\rho_1,\rho_2)|\leq 2\epsilon \right] \geq 1-2\delta.
\end{equation}

We begin by introducing $Y = \frac{1}{\ell} \sum_i X(\mathbf{k}_i)$, where $Y$ is an unbiased estimator of $F(\rho_1,\rho_2)$.
Therefore by Chebyshev's inequality we can say
\begin{equation}
\begin{split}
	\mathrm{Pr}\left[ |Y - F(\rho_1,\rho_2)| \geq \epsilon \right] &\leq \frac{1}{\epsilon^2 \ell} \\
	&\leq \delta
	\end{split}
\end{equation}
from the definition of $\ell$ in \Sec{FidelityEstimation}.

Now that we have a relation for $Y$ and the fidelity, we want to consider $\tilde{Y}$.
So far the procedure has followed previous work in \Refs{FlammiaLiu,Kliesch2020}, this is where a little more care is needed for the \su{d} algebra.
Substituting \Eq{XEstimatorFidelity} into \Eq{YEstimatorFidelity}
we yield
\begin{equation}\label{YEstimator2}
	\tilde{Y}= \frac{1}{\ell} \sum_{i=1}^{\ell}\sum_{j=1}^{m_i}  \frac{1}{m_i \mathcal{N}(\mathbf{k}_i)\,\cf(\mathbf{k}_i)}M_{ij},
\end{equation}
where $E[\tilde{Y}] = Y$.

For the multi-qubit case, we then move onto Hoeffding's inequality, where $M_{ij} \in  \{-1,1\}$.
This results in the bounds 
\begin{equation}
	b_{\pm} = \pm \frac{1}{m_i \mathcal{N}(\mathbf{k}_i)\,\cf(\mathbf{k}_i)}
\end{equation}
and more importantly to the denominator in Hoeffding's inequality
\begin{equation}
	|b| = b_{+} -b_{-}  =  \frac{2}{m_i \mathcal{N}(\mathbf{k}_i)\,\cf(\mathbf{k}_i)}.
\end{equation}
Applying Hoeffding's inequality
\begin{equation}\label{HoeffdingResult}
\begin{split}
	\mathrm{Pr}\left[ |\tilde{Y} - Y| \geq \epsilon  \right]& \leq 2\exp\left[ \frac{-\epsilon^2\ell^2}{\sum_{i=1}^\ell \frac{2}{m_i \mathcal{N}(\mathbf{k}_i)\,\cf(\mathbf{k}_i)}} \right] \\
	&\leq \delta
\end{split}
\end{equation}
given the definition for $m_i$.
By applying union bound this results in the desired
\begin{equation}
	\mathrm{Pr}\left[|\tilde{Y} - F(\rho_1,\rho_2)|\leq 2\epsilon \right] \geq 1-2\delta.
\end{equation}

Returning to \Eq{YEstimator2}, we note that when using the \su{d} algebra the results of $M_{ij}$ change.
Any single measurement of a state with any of the off-diagonal operator $\lambda^d_k$ will still yield the result $M \in  \{-1,1\}$.
This leaves the inequality in \Eq{HoeffdingResult} unchanged.
A measurement with any of the diagonal operators will on the other hand yield 
\begin{equation}
	M_{ij} \in  \left\{-\sqrt{\tfrac{2(n-1)}{n}},\sqrt{\tfrac{2}{n(n-1)}}\right\}.
\end{equation}
And so for these measurements, $\tilde{Y}$ is bounded by this new interval, dependent on the measurement choice.

Considering bipartite systems, we can choose two values, $n_1$ and $n_2$, where we say $n_1$ the choice of a diagonal operator for the first qudit and $n_2$ is the choice for the second qudit.
The measurement outcomes can now be anywhere between
\begin{equation}
	\left\{-\sqrt{\tfrac{2(n_1-1)}{n_1}}\sqrt{\tfrac{2}{n_2(n_2-1)}},\sqrt{\tfrac{2}{n_1(n_1-1)}}\sqrt{\tfrac{2}{n_2(n_2-1)}}\right\},
\end{equation}
resulting in the difference
\begin{equation}
	2\sqrt{\frac{n_1}{(n_1-1)n_2(n_2-1)}} \leq 2,
\end{equation}
since  $0<1/(n_2(n_2-1))<1/\sqrt{2}$, and $\sqrt{2n_1}/(n_1-1)\leq 2$.

The bound to put into Hoeffding's inequality are now
\begin{equation}
\begin{split}
		|b_{n_1 n_2}| &=  \frac{2}{m_i \mathcal{N}(\mathbf{k}_i)\,\cf(\mathbf{k}_i)}\sqrt{\frac{n_1}{(n_1-1)n_2(n_2-1)}}\\
	&\leq  \frac{2}{m_i \mathcal{N}(\mathbf{k}_i)\,\cf(\mathbf{k}_i)}
\end{split}
\end{equation}
so
\begin{equation}
\begin{split}
	\mathrm{Pr}\left[ |\tilde{Y} - Y| \geq \epsilon  \right] & \leq 2 \exp \left[ \frac{-\epsilon^2\ell^2}{\sum_{i=1}^\ell |b_{n_1 n_2}|} \right] \\
	&\leq 2\exp\left[ \frac{-\epsilon^2\ell^2}{\sum_{i=1}^\ell |b|} \right]\\
    &\leq \delta.
\end{split}
\end{equation}
Therefore, the results from \Sec{FidelityEstimation} hold for this choice of basis.

\subsection{Weyl algebra}

The Weyl algebra can be constructed from the generators $Z_d$ and $X_d$, defined
\begin{equation}\label{WeylAlgebraGenerators}
\begin{split}
	Z_d\ket{k}_d &= \ue{2\ui\pi/d} \ket{k}_d,\\
	X_d\ket{k}_d &= \ket{k \oplus 1}_d,
\end{split}
\end{equation}
where the $\oplus$ is addition modulo $k$.
The particulars of the construction of the characteristic function now depend on the dimension of the system in question, and there are multiple rules one can follow.
Much work has focussed on a Galois field theory construction that valid only for $d$ being prime, see for example \Refe{VOURDAS1997367}.
The prime construction has an advantage in that a set of $d+1$ mutually orthogonal bases (MUBs) exist for such dimensions.

Here we will focus on the construction from Gross in \Refe{Gross2016}, here we are concerned with $d$ being odd.
Note that this is informationally complete for any dimension as long as we only consider the characteristic function. 
\Refe{Gross2016} focussed on generating a discrete Wigner function for prime-dimensional systems through a discrete Fourier transform from the characteristic function; although this only defines a prime-dimensional Wigner function, the characteristic function can define a system of any dimension.

Given the generators in \Eq{WeylAlgebraGenerators}, the Weyl algebra can by fully described by the Weyl operator
\begin{equation}
	\mathcal{D}_d(p,q) = \exp(-\ui\pi pq/d) Z^pX^q.
\end{equation}
The characteristic function for one qudit can then be generated by taking
\begin{equation}
	\cf_\rho(p,q) = \frac{1}{\sqrt{d}}\Trace{\rho\;\mathcal{D}_d(p,q)^\dagger}.
\end{equation}
This is an informationally complete function, where the state can be recovered 
\begin{equation}
	\rho = \frac{1}{\sqrt{d}} \sum_{q,p}  \cf_\rho(p,q)\mathcal{D}_d(p,q).
\end{equation}

The fidelity can then be calculated 
\begin{equation}
	F(\rho_1,\rho_2) = \Trace{\rho_1\rho_2} = \sum \chi_{\rho_1}\overline{\chi}_{\rho_2}
\end{equation}
Note that in this case that the operators are now non-Hermitian and the characteristic function may have complex values.
Of course, since the $\mathcal{D}_d(p,q)$ operators are not observables, in practice the actual measurement of the characteristic function will be constructed from eigenstates of the given operator.
This is the advantage of choosing $p$-prime for this method, as we are guaranteed to have a set of $d+1$ MUBs~\cite{Wootters1989}.

Note that we also need to consider complex numbers.
Because of this, \Eq{XEstimatorFidelity} may need extra care, which gives using the \su{d} a preferential advantage.
However, The advantage of using the Weyl algebra is that $\mathcal{N}(k_i) = \sqrt{d}$ for all $i$, resulting in $\mathcal{N}(\mathbf{k}_i) = d$ in \Eq{XEstimatorFidelity}.

\clearpage
\section{Generalized Euler Angles and the construction of the two-qutrit strategy}\label{EulerAngles}
We begin by defining the generalized \SU{d} Euler angles, that can be created from a subset of the generators of \su{d}, that can be found in \App{sudAlgebra}.
The definition of the Euler angles presented here will differ from more standard approaches, where we will concentrate only on the generalization of a $\Sy$ rotations.
Where we define
\begin{equation}
	\mathsf{Y}^d_{k,1} = \ui \ket k \bra{0} - \ui \ket{0} \bra k
\end{equation}
Note that we only require this subset of the operators for the states we want to generate.

The rotation operator is then defined
\begin{equation}
	\mathcal{R}_d(\boldsymbol{\theta}) = \prod_{q=d}^2\prod_{p=2}^q \exp((-1)^{1-\delta_{q,d-1}}\ui\mathsf{Y}^d_{p,1}\theta_{p-1+j(q)}),
\end{equation}
where $j(q)=0$ when $q=d$ and $j(q)=\sum_{i=1}^{d-q}(d-i)$ otherwise. 
Applying this to the highest-weighted state then yields the coherent state
\begin{eqnarray}
	\ket{\boldsymbol{\theta}}_d &=& \mathcal{R}_d(\boldsymbol{\theta})\ket{d-1},
\end{eqnarray}
where 
\begin{equation}
	\langle k\ket{\boldsymbol{\theta}}_d = \begin{cases}
	\sin\theta_{d-1}\prod_{i=1}^{d-1}\cos\theta_{i}, & \text{if k = 0},\\
	\cos(\theta_{d-1})& \text{if  $k = d-1$},\\
	\sin\theta_{d-1}\sin\theta_{k}\prod_{i=k}^{d-1}\cos\theta_{i}, & \text{otherwise}.
	\end{cases}
\end{equation}
Note that only the first $d-1$ degrees of freedom have any effect on the highest-weighted state. 
This is ideal as this is the number of degrees of freedom needed to fully define the elements of the target state in the Schmidt decomposition.
The other degrees of freedom will, however, have an effect on the orthogonal states, this allows us to optimize over these extra degrees of freedom to find a more efficient strategy.

The elements of the coherent state then coincide with the elements of the target state in the Schmidt decomposition, \ket{\psi}, when we choose
\begin{equation}
	\theta_{k} = \begin{cases}
		\arctan\left(\frac{\langle 2 \ket{\psi}}{\langle 1 \ket{\psi}}\right) & \text{if $k = 1$},\\
		\arccos(\langle d-1 \ket{\psi}) & \text{if $k = d-1$},\\
		\arcsin\left(\frac{\langle k \ket{\psi}}{N(k,d)}\right)& \text{otherwise},
	\end{cases} 
\end{equation}
where
\begin{equation}
N(k,d) = \prod_{i=k+1}^{d-1} \sin(\theta_i).	
\end{equation}

We can now redefine the target state in terms of this coherent state structure, where
\begin{equation}
	\ket{\psi} = \sum_{k=0}^{d-1} \IP{k}{\boldsymbol{{\theta}}}\ket{kk}.
\end{equation}
An orthogonal basis can then be constructed by
\begin{equation}
	\ket{\psi^\perp_i} = \sum_{k=0}^{d-1} \bra{k}\mathcal{R}_d(\boldsymbol{\theta})\ket{i-1} \ket{kk},
\end{equation}
for $i=1...d-1$.

To generate separable states that are orthogonal to the target state, we then define
\begin{equation}\label{CreateGenSepStates}
\begin{split}
	\ket{\rho_i^\perp(\boldsymbol{\theta},\boldsymbol{\varphi})} &= \frac{1}{\mathcal{N}_i}\sum_{k=0}^{d-1} \ue{\ui\varphi_k}\sqrt{\bra{k}\mathcal{R}_d(\boldsymbol{\theta})\ket{i-1}} \ket{k}\\
	\ket{\sigma_i^\perp(\boldsymbol{\theta},\boldsymbol{\varphi})} &= \frac{1}{\mathcal{N}_i}\sum_{k=0}^{d-1} \ue{-\ui\varphi_k}\sqrt{\bra{k}\mathcal{R}_d(\boldsymbol{\theta})\ket{i-1}} \ket{k},
\end{split}
\end{equation}
where $\varphi_0=0$ and
\begin{equation}
	\mathcal{N}_i = \sqrt{\sum_{k=0}^{d-1} \ue{\ui\varphi_k}|\bra{k}\mathcal{R}_d(\boldsymbol{\theta})\ket{i-1}|}
\end{equation}

The general separable orthogonal states are then generated by
\begin{equation}
	\ket{\phi_i(\boldsymbol{\theta},\boldsymbol{\varphi})} = \ket{\rho_i^\perp(\boldsymbol{\theta},\boldsymbol{\varphi})} \otimes \ket{\sigma_i^\perp(\boldsymbol{\theta},\boldsymbol{\varphi})}.
\end{equation}

\clearpage
\begin{widetext}
\section{Elements of the qutrit strategy}\label{QutritStatElements}

To create general separable states that are orthogonal to \Eq{twoQutritState}, we need  states that are proportional to \Eq{twoQutritOrth1} and \Eq{twoQutritOrth2} in the relevant entries.
We then follow the procedure laid out in \App{EulerAngles} for $d=3$.
This results in the rotation operator
\begin{equation}
	\mathcal{R}_3(\boldsymbol{\theta}) =
	\begin{pmatrix}
		\cos\theta_1\cos\theta_2\cos\theta_3 - \sin\theta_1\sin\theta_3 && \cos\theta_1\cos\theta_2\sin\theta_3 + \sin\theta_1\cos\theta_3 && \cos\theta_1\sin\theta_2 \\
		\sin\theta_1\cos\theta_2\cos\theta_3 + \cos\theta_1\sin\theta_3 && \sin\theta_1\cos\theta_2\sin\theta_3 - \cos\theta_1\cos\theta_3 && \sin\theta_1\sin\theta_2 \\
		-\sin\theta_2\cos\theta_3 && -\sin\theta_2\sin\theta_3 && \cos\theta_2
	\end{pmatrix}.
\end{equation}
We can then create the coherent state
\begin{equation}
	\ket{\boldsymbol{\theta}}_3 = \mathcal{R}_3\ket{2} = \begin{pmatrix}
		\cos\theta_1\sin\theta_2 \\ \sin\theta_1\sin\theta_2 \\ \cos\theta_2
	\end{pmatrix}
\end{equation}
whose entries correspond to the coefficients to the Schmidt decomposition of a two-qutrit state, given in \Eq{twoQutritTargetState}.
We also have two orthogonal states
\begin{equation}
	\ket{\boldsymbol{\theta}^{\perp_1}}_3 = \mathcal{R}_3\ket{0} = \begin{pmatrix}
		\cos\theta_1\cos\theta_2\cos\theta_3 - \sin\theta_1\sin\theta_3 \\ \sin\theta_1\cos\theta_2\cos\theta_3 + \cos\theta_1\sin\theta_3  \\ -\sin\theta_2\cos\theta_3
	\end{pmatrix}
\end{equation}
and
\begin{equation}
	\ket{\boldsymbol{\theta}^{\perp_2}}_3 = \mathcal{R}_3\ket{1} = \begin{pmatrix}
		\cos\theta_1\cos\theta_2\sin\theta_3 + \sin\theta_1\cos\theta_3 \\ \sin\theta_1\cos\theta_2\sin\theta_3 - \cos\theta_1\cos\theta_3  \\ -\sin\theta_2\sin\theta_3.
	\end{pmatrix}
\end{equation}

From these, using \Eq{CreateGenSepStates}, we can then construct two separable states $\ket{\rho_7(\boldsymbol{\theta},\boldsymbol{\varphi})}\otimes\ket{\sigma_7(\boldsymbol{\theta},\boldsymbol{\varphi})}$ and $\ket{\rho_7^\perp(\boldsymbol{\theta},\boldsymbol{\varphi})}\otimes\ket{\sigma_7^\perp(\boldsymbol{\theta},\boldsymbol{\varphi})}$, where
\begin{eqnarray}
	\ket{\rho_7(\boldsymbol{\theta},\boldsymbol{\varphi})} &=& \frac{1}{\mathcal{N}_\rho^1} \Big( \sqrt{\cos\theta_1\cos\theta_2\cos\theta_3 - \sin\theta_1\sin\theta_3} \ket{0}  +\ue{\ui\varphi_1} \sqrt{\sin\theta_1\cos\theta_2\cos\theta_3 + \cos\theta_1\sin\theta_3} \ket{1}  \nonumber \\
	& &\hspace{7cm} + \;\ue{\ui\varphi_2}\sqrt{-\sin\theta_2\cos\theta_3} \ket{2} \Big)  \\
	\ket{\sigma_7(\boldsymbol{\theta},\boldsymbol{\varphi})} &=& \frac{1}{\mathcal{N}_\rho^1} \Big( \sqrt{\cos\theta_1\cos\theta_2\cos\theta_3 - \sin\theta_1\sin\theta_3} \ket{0} + \ue{-\ui\varphi_1} \sqrt{\sin\theta_1\cos\theta_2\cos\theta_3 + \cos\theta_1\sin\theta_3} \ket{1}  \nonumber \\
	& &\hspace{7cm} + \;\ue{-\ui\varphi_2}\sqrt{-\sin\theta_2\cos\theta_3} \ket{2} \Big) \\
	\ket{\rho_7^{\perp}(\boldsymbol{\theta},\boldsymbol{\varphi})} &=& \frac{1}{\mathcal{N}_\rho^1} \Big( \sqrt{\cos\theta_1\cos\theta_2\sin\theta_3 + \sin\theta_1\cos\theta_3} \ket{0} + \ue{\ui\varphi_1} \sqrt{\sin\theta_1\cos\theta_2\sin\theta_3 - \cos\theta_1\cos\theta_3} \ket{1}  \nonumber\\
	& &\hspace{7cm} + \;\ue{\ui\varphi_2}\sqrt{-\sin\theta_2\sin\theta_3} \ket{2} \Big) \\
	\ket{\sigma_7^{\perp}(\boldsymbol{\theta},\boldsymbol{\varphi})} &=& \frac{1}{\mathcal{N}_\rho^1} \Big( \sqrt{\cos\theta_1\cos\theta_2\sin\theta_3 + \sin\theta_1\cos\theta_3} \ket{0}  + \ue{-\ui\varphi_1} \sqrt{\sin\theta_1\cos\theta_2\sin\theta_3 - \cos\theta_1\cos\theta_3} \ket{1}  \nonumber \\
	& &\hspace{7cm} + \;\ue{-\ui\varphi_2}\sqrt{-\sin\theta_2\sin\theta_3} \ket{2} \Big),
\end{eqnarray}

where the $\mathcal{N}$ are the appropriate normalizations for the states and $(\boldsymbol{\theta},\boldsymbol{\varphi}) = (\theta_1,\theta_2,\theta_3,\varphi_1,\varphi_2)$.
 These states can now be put into \Eq{genRank7Strat}, where $\rho_7^j = \Proj{\rho_7(\boldsymbol{\theta},\boldsymbol{\varphi})}$, and likewise for $\sigma_7^j$, $\rho_7^{j\perp}$ and $\sigma_7^{j\perp}$.

\end{widetext}
\end{document}